\documentclass[twocolumn]{aastex631}

\newcommand{\arii}{\hbox{[Ar$\,${\scriptsize II}]}}
\newcommand{\ariii}{\hbox{[Ar$\,${\scriptsize III}]}}

\newcommand{\oiii}{\hbox{[O$\,${\scriptsize III}]}}
\newcommand{\oiv}{\hbox{[O$\,${\scriptsize III}]}}

\newcommand{\nii}{\hbox{[N$\,${\scriptsize II}]}}

\newcommand{\naiii}{\hbox{[Na$\,${\scriptsize III}]}}
\newcommand{\naiv}{\hbox{[Na$\,${\scriptsize IV}]}}
\newcommand{\navi}{\hbox{[Na$\,${\scriptsize VI}]}}

\newcommand{\neii}{\hbox{[Ne$\,${\scriptsize II}]}}
\newcommand{\neiii}{\hbox{[Ne$\,${\scriptsize III}]}}

\newcommand{\nev}{\hbox{[Ne$\,${\scriptsize V}]}}
\newcommand{\nevi}{\hbox{[Ne$\,${\scriptsize VI}]}}

\newcommand{\feii}{\hbox{[Fe$\,${\scriptsize II}]}}

\newcommand{\fevii}{\hbox{[Fe$\,${\scriptsize VII}]}}

\newcommand{\siii}{\hbox{[S$\,${\scriptsize III}]}}
\newcommand{\siv}{\hbox{[S$\,${\scriptsize IV}]}}

\newcommand{\mgv}{\hbox{[Mg$\,${\scriptsize V}]}}

\newcommand{\kms}{km\,s$^{-1}$} 
\newcommand{\msun}{M$_{\odot}$} 
\newcommand{\um}{$\mu$m}

\newcommand{\ergs}{erg\,s$^{-1}$}
\newcommand{\ergscm}{erg\,s$^{-1}$\,cm$^{-2}$}

\newcommand\jwst{\emph{JWST}}
\newcommand\spitzer{\emph{Spitzer}}

\newcommand\ifsfit{\texttt{IFSFIT}}

\newcommand\qtdfit{\texttt{q3dfit}}
\newcommand\questfit{\texttt{questfit}}

\shorttitle{\jwst\ F11119+3257 F05189-2524}
\shortauthors{Seebeck et al.}

\begin{document}

\title{\jwst\ Discovery of High-Velocity Mid-Infrared Ionized Outflows in Ultraluminous Infrared Galaxies F11119+3257 and F05189-2524}

\author[0000-0002-4014-9067]{Jerome Seebeck}
\affiliation{Department of Astronomy, University of Maryland, College Park, MD 20742, USA}

\author[0000-0001-5894-4651]{Kylie Yui Dan}
\affiliation{Department of Astronomy, University of Maryland, College Park, MD 20742, USA}

\author[0000-0002-3158-6820]{Sylvain Veilleux}
\affiliation{Department of Astronomy, University of Maryland, College Park, MD 20742, USA}
\affiliation{Joint Space-Science Institute, Department of Astronomy, University of Maryland, College Park, MD 20742, USA}

\author[0000-0002-1608-7564]{David Rupke}
\affiliation{Department of Physics, Rhodes College, Memphis, TN 38112, USA}

\author[0000-0001-5285-8517]{Eduardo Gonzalez-Alfonso}
\affiliation{Universidad de Alcalá, Departamento de Física y Matemáticas, Campus Universitario, 28871 Alcalá de Henares, Madrid, Spain}

\author[0000-0002-9627-5281]{Ismael Garcia-Bernete}
%\affiliation{Astrophysics, University of Oxford, DWB, Keble Road, Oxford OX1 3RH, UK}
\affiliation{Centro de Astrobiolog\'ia (CAB), CSIC-INTA, Camino Bajo del 497 Castillo s/n, E-28692 Villanueva de la Ca\~nada, Madrid, Spain}

\author[0000-0003-3762-7344]{Weizhe Liu}
\affiliation{Department of Astronomy, Steward Observatory, University of Arizona, Tucson, AZ 85719, USA}

\author[0000-0003-0291-9582]{Dieter Lutz}
\affiliation{Max Planck Institute for Extraterrestrial Physics, Giessenbachstraße 1, 85748 Garching, Germany}

\author[0000-0001-8485-0325]{Marcio Melendez}
\affiliation{Space Telescope Science Institute, 3700 San Martin Drive, Baltimore, MD 21218, USA}

\author[0000-0002-4005-9619]{Miguel Pereira-Santaella}
\affiliation{Instituto de Física Fundamental (IFF), CSIC, Serrano 123, E-28006 Madrid, Spain}

\author[0000-0002-0018-3666]{Eckhard Sturm}
\affiliation{Max Planck Institute for Extraterrestrial Physics, Giessenbachstraße 1, 85748 Garching, Germany}

\author[0000-0002-6562-8654]{Francesco Tombesi}
\affiliation{Physics Department, Tor Vergata University of Rome, Via della Ricerca Scientifica 1, 00133 Rome, Italy}
\affiliation{INFN – Rome Tor Vergata, Via della Ricerca Scientifica 1, 00133 Rome, Italy}
\affiliation{NASA Goddard Space Flight Center, Code 662, Greenbelt, MD 20771, USA}



\begin{abstract}

Ultra-fast outflows (UFOs) are thought to be a driving mechanism of large-scale winds driven by active galactic nuclei, which cause significant galactic feedback through quenching star formation and regulating supermassive black hole growth. We present James Webb Space Telescope (\jwst) Mid-Infrared Instrument Medium-Resolution Spectrometer observations of two nearby ultraluminous infrared galaxies (ULIRGs), F11119+3257 and F05189-2524, with nuclear X-ray detected UFOs and kiloparsec-scale outflow. These galaxies show remarkably similar mid-infrared continuum and emission line features, notably including a high-velocity $v_{90}$ $\sim$ 4000 \kms\ outflow detected in highly ionized neon emission lines, e.g., \nevi. In F05189-2524, we see a slightly slower biconical outflow extending up to $\sim2$ kpc in the same neon emission lines. Both sources show evidence of AGN-driven radiative feedback through a deficit of rotational molecular hydrogen lines in the nuclear region, $<$1 kpc from the central quasar, but no clear evidence of any molecular gas entrained in the quasar-driven outflow. Energetic analysis shows that the warm ionized gas in both of these sources contributes minimally ($\sim0.1-5\%$) to the momentum outflow rate of these sources and leaves the conclusions of previous literature unchanged: the energetics of these sources are broadly consistent with a momentum-conserving outflow. 

\end{abstract}



                        \section{Introduction} 
\label{sec:intro}

UltraLuminous InfraRed Galaxies (ULIRGs) are a class of infrared (IR) bright ($L_{\mathrm{IR}} \ge 10^{12}$ $L_{\odot}$) extragalactic sources, spanning a wide range of types from H~II region-like to Seyfert 1 and 2, that were first discovered with the InfraRed Astronomical Satellite (IRAS)  in 1984 \citep{Hou1985}. The bulk of their IR emission is due to energy re-radiated by warm dust excited by one or both of the two power sources: starburst and active galactic nuclei (AGN). Evidence showing a link between ULIRGs and galactic mergers has created a compelling theoretical picture: gas-rich galaxies merge, fueling gas into the galactic center and igniting dust-enshrouded star formation and supermassive black hole accretion. These engines drive galactic winds which provide negative feedback through clearing dust and gas from the galaxy, limiting star formation and leaving an exposed optical quasar that ultimately becomes a red gas-poor elliptical galaxy (e.g., \citealt{San1988, Vei2009, Vei2009b, Hop2016, Hic2018}). 

Evidence for these powerful winds is vital for our understanding of how inside-out negative feedback mechanisms work and has been detected in a majority of ULIRGs at a wide variety of physical scales and gas phases (e.g., molecular \citet{Vei2013}, neutral atomic \citet{Rup2017}, warm ionized \citet{Rup2013}, and hot ionized \citet{Vei2014, Liu2019, Liu2022}). While many ULIRGs are dominated by starbursts \citep{Vei2009}, high-velocity AGN-driven winds have been detected in a smaller subsample of sources \citep{Gon2017, Rup2017, Dan2025}. This makes local AGN-dominated ULIRGs, which can be observed with high spatial resolution, an excellent test bench for the higher redshift universe where mergers and AGN are more prevalent and impactful \citep{Hop2008, Zak2016},
% luminous, fast-wind quasars are common \citep[e.g.][]{Zak2016}, 
but direct observational evidence of feedback is limited \citep[e.g.][]{Vei2020}. 
% Cosmological simulations have recently shown that 
These luminous quasars are predicted to have significant effects on their host galaxies, 
% \citep[e.g.][]{Vog2014, Ric2016, Vei2020}
including, but not limited to, galaxy morphology \citep[e.g.,][]{Dub2016, Cho2018}, the interstellar medium \citep[e.g.,][]{Hop2016, Dav2019}, the circumgalactic medium \citep[e.g.,][]{Tum2017}, supermassive black hole growth \citep[e.g.,][]{Vol2016, Hop2016}, and star formation quenching \citep{Zub2012, Pon2017}.

Some models predict that galactic-scale AGN-driven winds are powered by sub-parsec-scale ultra-fast outflows (UFOs) \citep{Zub2012, Wag2013, Kin2015, Gas2020, Cos2020, Lah2021}. These UFOs are initially accelerated through radiatively or magnetically driven mechanisms based on the accretion rate of the SMBH \citep[e.g.,][]{Gal2023} and have been observed in numerous sources \citep{Tom2010, Tom2014, Mat2023, Xri2025}. Energy from the UFO is then deposited into the nuclear ISM, shocking the nearby gas, and creating an expanding hot bubble. If this bubble is unable to cool efficiently, then it will continue to expand adiabatically in an energy-conserving phase and produce a high ratio of the outer momentum outflow to the inner momentum outflow, powered by the UFO ($\dot P_{outer}/\dot P_{inner} > 1$). If the energy from the hot wind is radiated away, then the resulting outflow is expected to be momentum-conserving ($\dot P_{outer}/\dot P_{inner} \lesssim 1$) and driven largely by ram pressure. There is no clear consensus on a dominant outflow coupling mechanism, and evidence for both energy-conserving and momentum-conserving outflows has been reported (e.g., \citealt{Vei2005}; \citealt{Tom2015}; \citealt{Smi2019}; \citealt{Vei2020}; \citealt{Mar2020}; \citealt{Lah2021}; \citealt{Lan2024}). 

% JWST Significance note

% Over the years, integral-field unit spectroscopy (IFS) has been an extremely useful tool in mapping galaxy-scale extended outflows from local quasars and other galaxies \citep{Nes2008, Rup2011, Liu2013a, Liu2013b, Sin2013, Car2015, Cre2015, Fin2017, Rup2019}. However, the main challenge in luminous AGN, even nearby ones, lies in separating the bright quasar light from the extended host galaxy emission. The excellent spatial resolution and stable point spread function (PSF) of JWST at near- and mid-infared (NIR and MIR) wavelengths have been a true game-changer for this research \citep[e.g.][]{Wyl2022, Vay2023, Vei2023, Rup2023a}. 

In this paper, we analyze the mid-infrared emission of two AGN-dominated, late-stage merger ULIRGs with X-ray detected UFOs, F11119+3257 and F05189-2524, in order to search for and parameterize outflow dynamics and other evidence of AGN feedback. In Section \ref{sec:obs_redux}, we discuss the observations and reduction of the \jwst\ data. We discuss our analysis methods, including the use of the integral-field unit spectroscopy (IFS) analysis software package \qtdfit, in Section \ref{sec:data_analysis}. In Section \ref{sec:results} we present the results of our analysis of both extracted spectra and IFS kinematic maps. In Section \ref{sec:discussion}, we discuss the evidence and energetics of the high-velocity outflow of both sources. We summarize our conclusions in Section \ref{sec:conclusion} 

For this paper, we adopt the flat $\Lambda$CDM cosmology: $H_0$ = 70 \kms\ Mpc$^{-1}$, $\Omega_m = 0.3$ and $\Omega_{\Lambda} = 0.7$. For F11119+3257, we use a redshift of $z$ = 0.19 $\pm$ 0.01, derived by \citet{Pan2019}. This gives a luminosity distance $D_L=$ 925.6 Mpc and a physical scale of 1\arcsec\ = 3.169 kpc. For F05189-2524, we use a redshift of $z$ = 0.042731, derived by \citet{Lam2022}. This gives a luminosity distance $D_L=$ 172.4 Mpc and a physical scale of 1\arcsec\ = 0.848 kpc. All emission lines are identified by their wavelengths in vacuum (e.g., \oiii\ $\lambda$5008 \AA).

\subsection{F11119+3257}
\label{subsec:f19_intro}

IRAS F11119+3257 is a relatively nearby ($z$ = 0.19) Seyfert 1 late-stage merger ULIRG and has a $L_{\mathrm{bol}}\sim10^{46.3}$ \ergs,  71\% of which is attributed to AGN emission \citep{Vei2009}. F11119+3257 was the first source to have confirmed observations of both galactic-scale and sub-parsec relativistic outflows \citep{Tom2015}. This was a spectacular link between central AGN-powered winds and galactic-scale feedback. The nuclear wind was detected through Suzaku Fe K absorption observations (later confirmed with NuSTAR \citealt{Tom2017}), which showed a $v_{out}\sim$ 0.3c outflow with a momentum outflow rate similar to that input by the central black hole. Recent XMM-Newton and NuSTAR observations \citep{Lan2024} have confirmed the UFO and estimated a slightly higher momentum outflow rate, $\sim6L_{\mathrm{bol}}/c$. Unresolved cold molecular outflows, traced by OH and modeled to a distance of 300 pc \citep{Tom2015}, imply that the energy has been conserved from this UFO to the galactic-scale outflow. Updated CO (1-0) observations from \citet{Vei2017} show an extended outflow not inconsistent with a momentum-conserving flow. Spatially resolved outflow has not been detected for any gas phase in this source. 

% SOURCE NOTES
% - basic stuff (redshift, seyfert 11, merger type, AGN contribution, BH mass, eddington ratio)
%      -   z     Lbol    type interaction  separation   AGN
%      - 0.189  12.67     S1    IVb          <3.16      .71
% - x-ray UFO observations (Tombesi 2015)
% - new x-ray XMM-Newtron (Lanzuisi1 24)
% - molecular outflows CO and OH (veilleux 2013, 2017)
% - HST optical measurements Lya and NaD (Liu 2022)
% - two sided jet in radio (yang 2020)

\subsection{F05189-2524}
\label{subsec:f05_intro}

IRAS F05189-2524 is a relatively nearby ($z$ = 0.042731) Seyfert 2 late-stage merger ULIRG with a broad line region observed in the near-infrared \citep{Vei1999}. It is an extremely X-ray bright source with a Fe K absorption feature blueshifted from E = 6.966 to 7.8 keV, implying an UFO with $v_{out}$ = 0.11c \citep{Smi2019}. This UFO was recently confirmed by XRISM with the detection of three separate Fe K absorption features in the 7.8 keV region, ranging from 0.76c $-$ 0.143c \citep{Nod2025}. F05189-2524 has a $L_{\mathrm{bol}}\sim10^{45.8}$ \ergs,  69\% of which is attributed to AGN emission \citep{Vei2009}. Previous observations have shown spatially resolved and unresolved high-velocity outflows in many phases of gas. Cold molecular outflows, traced by OH and CO \citep{Gon2017, Flu2019, Lut2020, Lam2022}, show a momentum outflow rate of $\sim 3\:L_{\mathrm{bol}}/c$. IFS observations of resolved outflows, R$_{\mathrm{out}}\le3$ kpc, with median velocities $\sim$ 500 \kms\ in neutral gas, traced by Na I D 5890, 5896 \AA, and warm ionized gas, traced by \nii\ 6585 \AA, show momentum boosts of $\sim 3\:L_{\mathrm{bol}}/c$ and $\sim 0.04\:L_{\mathrm{bol}}/c$, respectively \citep{Rup2017}.

% SOURCE NOTES
% - basic stuff (redshift, seyfert 11, merger type, AGN contribution, BH mass, eddington ratio)
%     -   z    Lbol    type interaction  separation      AGN
%     - 0.19   12.22    S2     IVb           <0.13       .69
% - x-ray UFO observations (Smith 2019, older suzuka obs Teng 2009)
% - OH outflow from (gonzalez alfronso 2017, veilleux 2013)
% - HI with GBT (Teng 2013)
% - spatially resolved NaDI and OIII GMOS (rupke 2015, 2017)
% - Optical Ha etc with VLT  (Westmoquette 2012, Bellocchi 2012)
% - Nuclear warm molecular gas (Pereira-Santaella 2014)
% - HST imaging (Veilleux 2006, or 2009) (2002 ground based)
% - radio???

\section{Observations and Data Reduction} 
\label{sec:obs_redux}

Both F11119+3257\footnote{The \jwst\ MIRI/MRS data used for F11119+3257 in this paper can be found in MAST: \dataset[10.17909/1bgd-mf94]{http://dx.doi.org/10.17909/1bgd-mf94}.} and F05189-2524\footnote{The \jwst\ MIRI/MRS data used for F11119+3257 in this paper can be found in MAST: \dataset[10.17909/5zxm-mq78]{http://dx.doi.org/10.17909/5zxm-mq78}.} were observed using the Medium-Resolution Spectrometer (MRS) mode of the Mid-InfraRed Instrument \citep[MIRI]{Wri2023, Arg2023}. All of the grating settings, Short, Medium, and Long, were used to achieve the full wavelength range of MIRI (4.9 to 27.9 $\mu m$). The spectral resolution of MIRI ranges from 8 \AA\ in channel 1 to 60 \AA\ in channel 4, corresponding to $30-85$ \kms\ across the entire MIRI wavelength range. A 4-point extended dither pattern was used to help remove background contamination and reduce undersampling. F11119+3257 was observed on May 23, 2024 (PID 3869, PI Veilleux) and F05189-2524 was observed on March 4, 2024 (PID 3368, PI Armus). 

We reduced these data with \jwst\ v1.15.1 \citep{Bus2022, Bus2024} and v12.0.2 of the Calibration Reference Data System (CRDS). The reduction is done with the MIRI pipeline sample notebook publicly available on the Space Telescope github (\href{https://github.com/STScI-MIRI/MRS-ExampleNB/blob/main/Flight_Notebook1/MRS_FlightNB1.ipynb}{MRS\_FlightNB1.ipynb}) with the 2D, pixel-by-pixel, background subtraction step turned on and the rest of the notebook in its default state. This step noticeably improves the overall noise in fully reduced cubes. 

\begin{figure}
    \centering
    \includegraphics[width=.9\columnwidth]{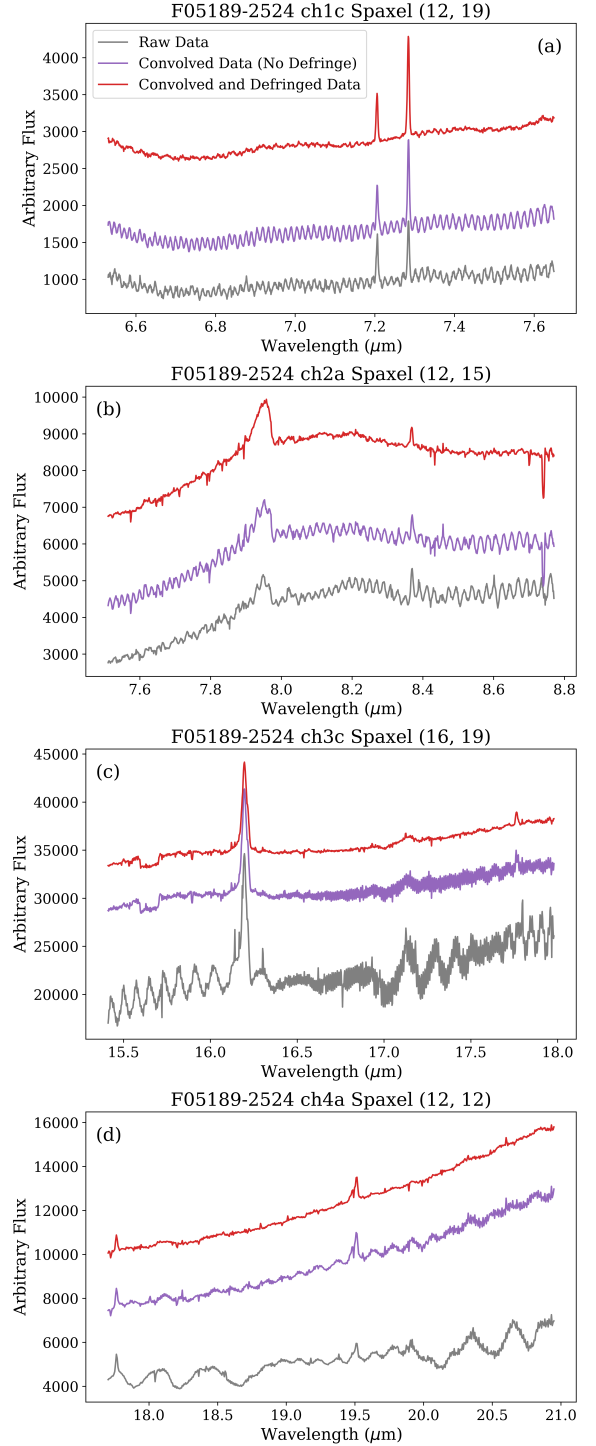}
    \caption{Spectra extracted from 4 sample spaxels, one for each MIRI channel, showing the post \jwst\ pipeline defringing process for the source F05189-2524. Each subplot shows the raw data directly from the \jwst\ pipeline (gray), the data smoothed with a running circular average using a radius of 1.5 spaxels (purple), and the final resulting spectra which has been smoothed and defringed using the \jwst\ pipeline function \texttt{fit\_residual\_fringes\_1d} (red). %In panel (a) we display spaxel (12, 19) from the long (C) band of channel 1. In panel (b) we display spaxel (12, 15) from the short (A) band of channel 2. In panel (c) we display spaxel (16, 19) from the long (C) band of channel 3. In panel (d) we display spaxel (12, 12) from the short (A) band of channel 4.}
    }
    \label{fig:defringing}
\end{figure}

We can see the results from the \jwst\ pipeline applied to F05189-2524 in Figure \ref{fig:defringing}, which displays spaxels from sub-bands in each MIRI channel as the gray spectra. Two types of fringing remain in these reduced cubes. The first can be seen in panels (c) and (d) of Figure \ref{fig:defringing}, showing large fringes at the edge of the MIRI bands but minimal fringing near the center. This is caused by spatial undersampling, which is an expected side effect of maximizing the field of view (FOV) and spectral resolution of \jwst\ \citep{Law2023}. The recommended solution for this issue is to extract spectra with a width similar to that of the PSF. 

The second type of fringing is seen across all 4 channels and increases with higher frequencies. To mitigate this, we take a two-step approach to correct each MRS sub-band individually. First, we smooth the cube with a running circular average using a radius of 1.5 pixels, seen as the purple spectra in Figure \ref{fig:defringing}. This smoothing eliminates the fringes caused by spatial undersampling and increases the S/N ratio of the higher frequency fringes. Second, we apply the \jwst\ pipeline function \texttt{fit\_residual\_fringes\_1d} spaxel-by-spaxel to the reduced cubes; the final sample spectra can be seen as the red spectra in Figure \ref{fig:defringing}. We find that the efficacy of this function is significantly improved by the previous smoothing step. In our final re-reduced cube, we find that both fringing types are eliminated or significantly reduced in a majority of spaxels, allowing for much more detailed analysis of line kinematics. 

% For spatial comparisons of spectrally distant emission lines, we spatially smooth a set of the final reduced cubes for each sub-band individually, unifying the effective PSF size to the observed wavelength of the H$_2$ S(1) 17.03 $\mu$m line. This is the emission line with the longest wavelength that we analyze with spatially resolved data and corresponds to a final PSF FWHM of 1.46\arcsec\ for F11119+3257 and 0.99\arcsec\ for F05189-2524. We estimate the wavelength-dependent PSF FWHM of MIRI MRS from the value observed in 3D drizzled cubes (Figure 12 of \citealt{Arg2023}). Then, at every wavelength, we calculate the quadrature difference ($\sigma_{convolve} = \sqrt{\sigma^2_{17.03}-\sigma_{\lambda}^2}$) to the PSF FWHM of the H$_2$ S(1) 17.03 $\mu$m line and convolve the slice with a 2D Gaussian with $\sigma=\sigma_{convolve}$. We only use these cubes when direct spaxel-to-spaxel comparisons need to be made and will defer to the higher spatial resolution of the non-unified cubes otherwise. We will refer to these cubes as the ``PSF-matched'' cubes. 

%  to mitigate issues caused by the wavelength dependent PSF size and MRS sub-band flux calibrations

For spatial comparisons of spectrally distant emission lines, we apply two more reduction steps. First, we spatially smooth a set of the final reduced cubes for each sub-band individually, unifying the effective PSF size to the observed wavelength of the H$_2$ S(1) 17.03 $\mu$m line. This is the emission line with the longest wavelength that we analyze with spatially resolved data and corresponds to a final PSF FWHM of 1.46\arcsec\ for F11119+3257 and 0.99\arcsec\ for F05189-2524. We estimate the wavelength-dependent PSF FWHM of MIRI MRS from the value observed in 3D drizzled cubes (Figure 12 of \citealt{Arg2023}). Then, at every wavelength, we calculate the quadrature difference ($\sigma_{convolve} = \sqrt{\sigma^2_{17.03}-\sigma_{\lambda}^2}$) to the PSF FWHM of the H$_2$ S(1) 17.03 $\mu$m line and convolve the slice with a 2D Gaussian with $\sigma=\sigma_{convolve}$. Second, we coarsely stitch the cubes together to remove the effects of any flux discontinuities in a similar method to that described in \citet{Cec2025}. To do this, we first rescale the flux down to the lowest spatial resolution channel using the flux-conserving method in the package \texttt{reproject} \citep{Rob2024}. For both sources, this is again limited by the longest wavelength observed line H$_2$ S(1) 17.03 \um. For F11119+3257, this limits the spatial resolution to that of Channel 3, 0.2\arcsec/pixel, and for F05189-2524, the resolution is limited to that of Channel 4, 0.35\arcsec/pixel. Then, we stitch the cubes together by scaling the emission of the bluer cube to match the flux of the redder cube in each spaxel, starting with the reddest channel used in the analysis of each source. This results in a continuous spectrum in each spaxel of the data cube across the entire MIRI range. We only use these cubes when direct spaxel-to-spaxel comparisons need to be made of emission lines not in the same sub-band and will defer to the higher spatial resolution of the non-unified cubes otherwise. We will refer to these cubes as the ``unified'' cubes.

\begin{figure*}
    \centering
    \includegraphics[width=\textwidth]{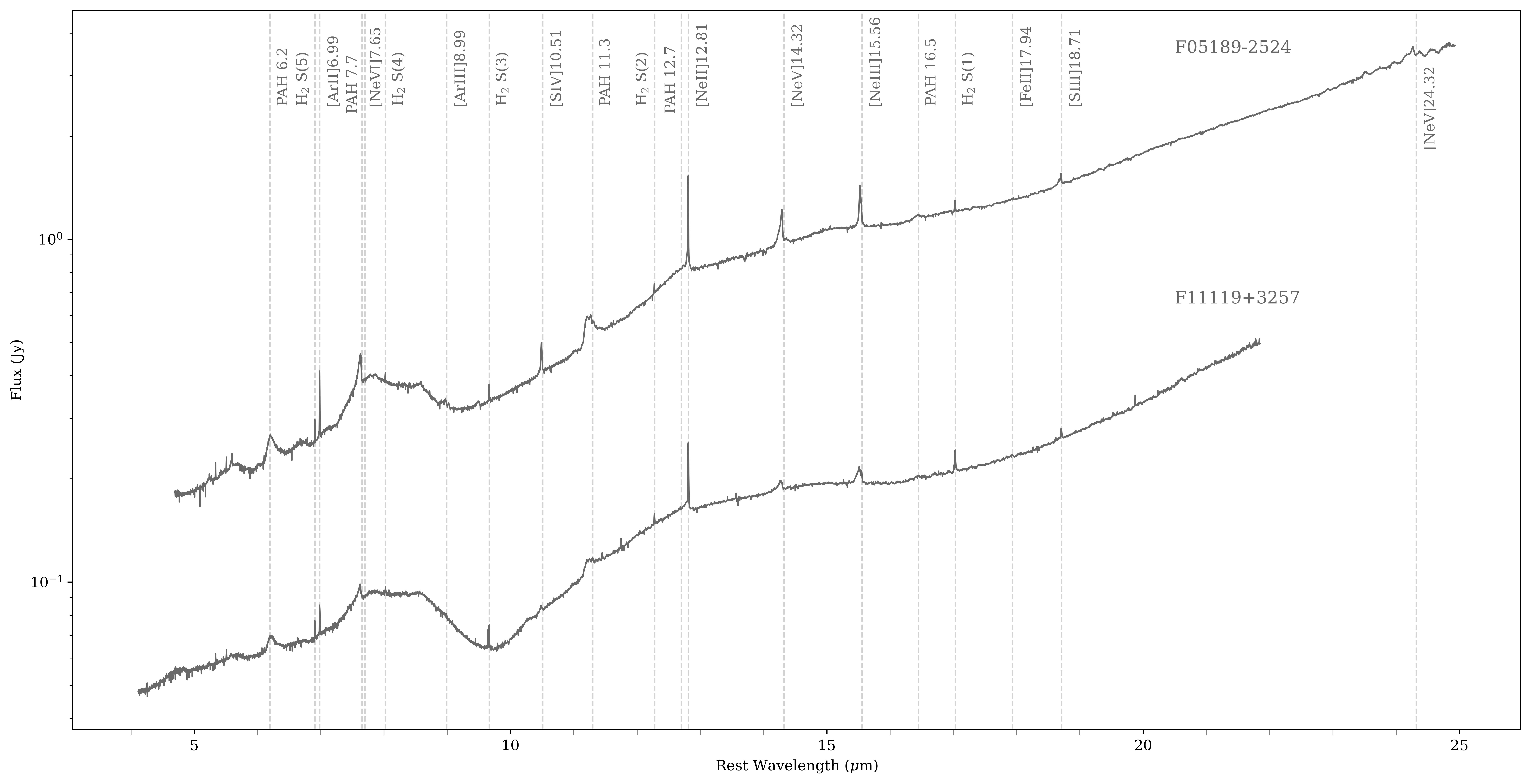}
    \caption{1D nuclear spectra of F05189-2524 (top) and F11119+3257 (bottom) extracted from the \jwst\ pipeline data cube and corrected for residual fringes. Common MIR spectral features are marked with vertical dashed lines at their rest wavelengths.}
    \label{fig:1dspec}
\end{figure*}

For the 1D spectral analysis, we follow a similar method to \cite{See2024}, extracting a nuclear spectrum for both sources out of data cubes directly from the pipeline using an aperture with a FWHM$(\lambda) = 0.7$\arcsec\ for $\lambda < 8$ \um\ and FWHM$(\lambda) = 0.7\arcsec \times \lambda[$\um$]/8$ \um\ for $\lambda \geq 8$ \um. Data from each channel are first defringed using the \texttt{fit\_residual\_fringes\_1d} function and then concatenated to form a longer spectrum. For overlapping regions, we take the average of both sub-bands which causes no spectral jumps as seen in the final extracted spectra in Figure \ref{fig:1dspec}. We have rebinned the spectrum on the grid with $\Delta\lambda$ = 8 \AA, the spectral element width of Channel 1. The rebinning causes no significant change ($\le$2\%) in the fluxes of longer wavelength lines but improves the flux measurements of shorter wavelength lines by up to 20\% from measurements using the coarser spectral width of Channel 4.

% Using the methods described in \cite{See2024} we then apply a wavelength-dependent correction, calibrated from the 1D spectra, to both the extracted 1D spectrum and 3D spectral data cubes of both sources to correct for the sensitivity drop at longer \jwst\ wavelengths noted by the \jwst\ MIRI MRS team\footnote{See \href{https://www.stsci.edu/contents/news/jwst/2023/performance-monitoring-reveals-a-decrease-in-the-miri-mrs-throughput-at-the-longest-wavelengths?page=5&filterUUID=0655d914-43ee-4d09-a91b-9f45be575098}{April 21, 2023 \jwst\ Observer News Article}}. The final extracted 1D spectra of both sources with common emission lines marked can be seen in Figure \ref{fig:1dspec}. 

\section{Data Analysis}
\label{sec:data_analysis}

\subsection{Nuclear Spectrum Fitting}

To fit the nuclear spectrum of both sources, we use two methods. First, we use the \qtdfit\ adaptation of \questfit\ \citep{Sch2008, Rup2021} to fit larger continuum features. While this method is effective in fitting broad continuum features, it is less reliable over narrow wavelength regions, causing the line fits to be less accurate. To make line flux measurements, we use the \qtdfit\ method \texttt{polyfit} to more effectively subtract the continuum from nearby emission lines. 

\questfit\ was built to fit mid-infrared (MIR) spectra of galaxies from the \spitzer\ Quasar and ULIRG Evolution Study \citep[QUEST;][]{Vei2009}. It uses polycyclic aromatic hydrocarbon (PAH) templates \citep{Smi2007}, silicate templates \citep{Sch2008, Gro2006}, extinction and absorption models, and blackbodies of various temperatures. In using \questfit\ for the fitting of our extracted nuclear spectra, we follow the same methods as \citet{See2024}, which follows a similar procedure to that described in \cite{Vei2009}. 

To fit PAH features, we subtract the non-PAH emission of the \questfit\ model from the extracted spectrum and directly fit the PAH residuals. We extract fluxes from the PAH 6.2, 7.7, 11.3, and 12.7 \um\ features by simultaneously fitting their main features with a sum of Lorentzian profiles superimposed on a continuum approximated by a second-order polynomial. This follows the methods described in \cite{See2024} and \cite{Sch2006} and should give similar values to those derived in \cite{Rig2021}.

\subsection{Integral Field Spectroscopy Fitting}
We use the software package \qtdfit\ for the majority of our analysis \citep{Rup2023}. This package is based on the IDL software \ifsfit\ \citep{Rup2014, Rup2017}, designed to remove the PSF caused by the central compact quasar emission, allowing us to see the much fainter emission from the host galaxy without contamination from the bright PSF. This tool is ideally suited for the analysis of IFS AGN data, as without removing a dominant PSF, the compact nuclear region overwhelms emission from extended regions of the galaxy. More details on the \qtdfit\ software and its use in the analysis of \jwst\ IFS data can be found in \citet{Vei2023b, Rup2023a, Vay2023}. 

In our fits, we use a spectrum extracted from the brightest spaxel in the data as a quasar template. In an initial fit, the template, in combination with emission lines and an exponential starlight model, is fit to the data. The template is scaled up or down with a series of exponentials to match the continuum level of different spaxels and remove data that resembles the nuclear spectrum. These initial fits may then be used as initial guesses in a second stage of fitting where more detailed models, such as stellar population synthesis, can be used to lower the residuals. The process then loops, refitting emission lines and the total sum continuum until certain residual levels in the relative error of the approximate solution and desired sum of squares are met. Line profiles are fit with a specified number of Gaussian components to the spectrum with starlight and nuclear emission removed. \qtdfit\ also imposes a significance cut on each component of each line every iteration. If the fit is not significant enough, it will be removed and fit with fewer components or not fit at all. This entire process is done for every spaxel in the data and can be accelerated with multicore processing. Due to a lack of obvious stellar absorption features in the data, the use of exponential starlight models was only used sparingly in certain spaxels to improve the continuum fit.

To reduce computational times and improve fit quality, we run all of our \qtdfit\ analysis in the sub-band where each line is observed. \qtdfit\ relies on strong initial guesses for the central velocity and width of Gaussian line components. When two non kinematically tied lines with very distinct kinematic components are present in the same sub-band, even the perfect initial guess can cause false-positive or negative results in the fits. To mitigate this problem, we occasionally fit the sub-band separately for the different lines. 

\section{Results}
\label{sec:results}

\subsection{Nuclear Emission}
\label{subsec:res_nuclear}

\begin{figure}
    \centering
    \includegraphics[width=\columnwidth]{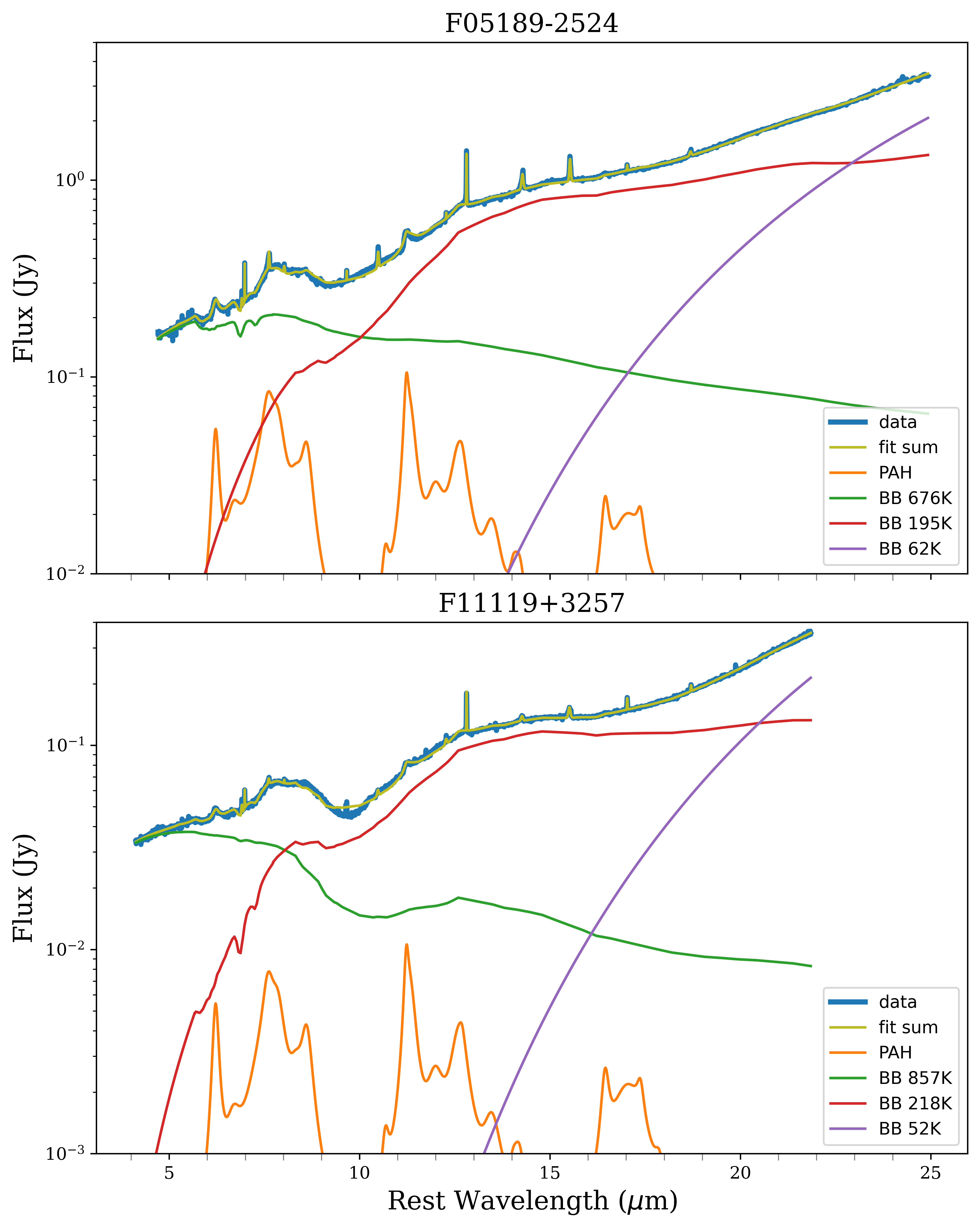}
    \caption{Fit of 1D extracted spectrum (blue) with \questfit\ the continuum fitting package within \qtdfit\ for F05189-2524 (top) and F11119+3257 (bottom). Each spectrum is fit with three black bodies with different temperatures, cool (purple), warm (red), and hot (green), and a PAH template (orange). The total sum fit is the yellow curve. }
    \label{fig:nuc_quest}
\end{figure}

Our fitting methods well approximate the continuum of both sources with slight fitting residuals near the $\tau_{9.7}$ silicate absorption and some emission lines. These fits are shown in Figure \ref{fig:nuc_quest}. Our PAH measurements are listed in Table \ref{tab:PAH}. Due to minor continuum fitting issues in F11119+3257, we are only able to set upper limits on the 12.7 \um\ PAH feature based on the fit PAH template instead of directly fitting the residuals as in the other features. We find our 6.2 and 7.7 \um\ feature measurements consistent, within error bars and expected instrumental aperture differences, with results from \spitzer\ data in \citet{Vei2009}. There are no previous measurements of the 11.3 and 12.7 \um\ PAH features. Our results show that the equivalent widths (EW = $F$(PAH)/$F_{\lambda,cont}(\lambda_{\mathrm{PAH}})$) of PAH features are stronger in F05189-2524 than in F11119+3257, but the overall PAH luminosities are larger in F11119+3257 than in F05189-2524.  

\begin{deluxetable*}{l c c c c c}
 \tablecaption{PAH Measurements
 \label{tab:PAH}}
 \tablehead{\colhead{PAH Feature} & \colhead{Source} & \colhead{6.2 \um} & \colhead{7.7 \um} & \colhead{11.3 \um} & \colhead{12.7 \um}}
 \startdata
    $\lambda$ range (\um) & & 5.9 $-$ 6.5 & 6.9 $-$ 9.2 & 10.8 $-$ 11.7 & 12.5 $-$ 12.8\\
    EW\tablenotemark{a}& F11119+3257 & 2.6 (0.5) & 7.3 (4.0) & 1.4 (0.8) & $\le$2.1\tablenotemark{*}\\
     & F05189-2524 & 4.6 (1.4) & 25.8 (7.8) & 3.4 (0.3) & 2.4 (1.1)\\
    Flux\tablenotemark{b} & F11119+3257 & 0.9 (0.1) & 2.2 (1.1) & 0.3 (0.1) & $\le$0.4\tablenotemark{*}\\
     & F05189-2524 & 9.0 (1.0) & 36.2 (11.9) & 3.6 (0.3) & 3.2 (1.4)\\
    Luminosity\tablenotemark{c} & F11119+3257 & 2.3 (0.1) & 5.7 (3.0) & 0.7 (0.2) & $\le$1.2\tablenotemark{*}\\
     & F05189-2524 & 1.0 (0.2) & 4.1 (1.2) & 0.4 (0.1) & 0.4 (0.2)\\
 \enddata
 \tablenotetext{a}{In units of $10^{-2}$ \um}
 \tablenotetext{b}{In units of $10^{-13}$ \ergscm}
 \tablenotetext{c}{In units of 10$^9$ $L_\odot$}
 \tablenotetext{*}{Upper limits derived from PAH template fits}
 \tablecomments{Errors for our PAH fits are listed (in parentheses) and are taken from the standard error of the least squares fit to the data.}
\end{deluxetable*}

We fit 19 emission lines (11 ionized and 8 molecular) in F11119+3257 and 25 emission lines (17 ionized and 8 molecular) in F05189-2524.  For both sources, emission lines are fit with one component except \neiii, \nev, and \nevi\ (3 components) and \mgv, \ariii, and \neii\ (2 components). The $v_{50}$ and $w_{80}$ of the ionized gas fit is shown in Figure \ref{fig:IPvsv50} plotted against the ionization potential (IP) and critical density (gathered from PyNeb, \citealt{Lur2015}) of the gas. $v_{50}$ is the 50-percentile or median velocity: the velocity where 50\% of the line flux is accumulated as calculated from the red side of the profile. $w_{80}$ is the 80-percentile velocity width: the velocity width of the line containing 80\% of line flux, centered on $v_{50}$. See \cite{Vei1991c, Vei1991a, Vei1991b, Zak2014} for a more detailed description and validation of these measurements. The fluxes, $v_{50}$, and $w_{80}$ from all line fits, for both sources are shown in Table \ref{tab:nuc_res}. Both sources show a significant increase in both the $|v_{50}|$ and $w_{80}$ with increases in IP but no significant trend for the critical density. F11119+3257 shows a steeper correlation between outflow velocities and IP than F05189-2524. The high velocities, IPs, and positive correlation between the two are possibly indicative of a decelerating AGN-driven outflow, first proposed in the MIR by \citet{Spo2009}. Another possible explanation for this trend is that the outflow is spatially stratified with larger velocities and IP emission closer to the outflow axis and slower velocities towards the edges, as supported by simulations \citep[e.g.][and references therein]{Sch2024}.

\begin{deluxetable*}{l c c c c c c c c }
 \tablecaption{Nuclear Results
 \label{tab:nuc_res}}
 \tablehead{\colhead{} & \colhead{} & \colhead{} & \multicolumn{3}{c}{\centering F05189-2524} & \multicolumn{3}{c}{\centering F11119+3257} \\
  \cline{4-6} \cline{7-9} 
 \colhead{Feature ID} & \colhead{$\lambda_{rest}$} & \colhead{IP}& \colhead{Flux} & \colhead{$v_{50}$} & \colhead{$w_{80}$} & \colhead{Flux} & \colhead{$v_{50}$} & \colhead{$w_{80}$} \\
 \colhead{} & \colhead{$\mu$m} & \colhead{eV}& \colhead{$10^{-14}$ \ergscm} & \colhead{\kms} & \colhead{\kms} & \colhead{$10^{-14}$ \ergscm} & \colhead{\kms} & \colhead{\kms} }
 \startdata
   H$_2$ $0-0$ S(8) & 5.05 &  --- & 0.3 (0.2) & -130 (10) & 310 (103) & 0.1 (0.1) & -90 (10) & 270 (54) \\
   \feii & 5.34 & 7.9 & 0.9 (0.1) & -100 (10) & 260 (20) & 0.2 (0.1) & -70 (10) & 320 (23) \\
   H$_2$ $0-0$ S(7) & 5.51 &  --- & 0.8 (0.1) & -120 (10) & 230 (20) & 0.2 (0.1) & -40 (10) & 300 (21) \\
   \mgv & 5.61 & 109.3 & 2.6 (0.7) & -870 (10) & 1040 (145) & 0.2 (0.1) & -1430 (10) & 1340 (218) \\
   H$_2$ $0-0$ S(6) & 6.11 &  --- & 0.4 (0.1) & -100 (10) & 260 (20) & 0.1 (0.1) & -20 (10) & 560 (83) \\
   \nii & 6.64 &  14.5 & 0.5 (0.2) & -10 (10) & 730 (144) & 0.1 (0.1) & 0 (10) & 620 (70) \\
   H$_2$ $0-0$ S(5) & 6.91 &  --- & 1.3 (0.7) & -90 (10) & 240 (66) & 0.3 (0.1) & -70 (10) & 270 (20) \\
   \arii & 6.99 & 15.8 & 5.1 (0.2) & -90 (10) & 270 (20) & 0.6 (0.1) & -60 (10) & 330 (20) \\
   \naiii & 7.32 &  47.3 & 0.8 (0.2) & -600 (10) & 680 (75) & 0.1 (0.1) & -740 (10) & 1100 (257) \\
   \nevi & 7.65 & 126.2 & 13.5 (3.0) & -1310 (17) & 1670 (188) & 1.7 (0.3) & -1770 (10) & 2670 (253) \\
   H$_2$ $0-0$ S(4) & 8.03 &  --- & 0.7 (0.3) & -120 (10) & 270 (51) & 0.2 (0.1) & -70 (10) & 560 (56) \\
   \navi & 8.61 & 138.4 & 0.6 (0.2) & -1000 (10) & 850 (110) &  --- & --- & --- \\
   \ariii & 8.99 & 27.6 & 3.2 (0.5) & -1020 (10) & 2210 (183) &  --- & --- & --- \\
   \naiv & 9.04 & 71.6 & 0.3 (0.3) & -680 (10) & 540 (217) &  --- & --- & --- \\
   \fevii & 9.53 & 99.0 & 1.2 (0.3) & -1450 (10) & 1520 (221) &  --- & --- & --- \\
   H$_2$ $0-0$ S(3) & 9.66 &  --- & 1.0 (0.1) & -100 (10) & 250 (20) & 0.3 (0.1) & -40 (10) & 310 (20) \\
   \siv & 10.51 & 34.9 & 5.0 (0.2) & -630 (10) & 680 (20) & 0.2 (0.1) & -820 (10) & 1060 (58) \\
   H$_2$ $0-0$ S(2) & 12.28 &  --- & 0.8 (0.1) & -100 (10) & 220 (20) & 0.2 (0.1) & -80 (10) & 280 (20) \\
   \neii & 12.81 & 21.6 & 18.8 (0.6) & -130 (10) & 590 (20) & 2.5 (0.1) & -110 (10) & 950 (26) \\
   \nev & 14.32 & 97.2 & 17.1 (0.9) & -920 (10) & 1520 (41) & 0.7 (0.1) & -1220 (10) & 1440 (117) \\
   \neiii & 15.56 & 41.0 & 16.3 (0.6) & -560 (10) & 980 (20) & 1.6 (0.1) & -900 (10) & 1680 (63) \\
   H$_2$ $0-0$ S(1) & 17.03 &  --- & 1.3 (0.1) & -140 (10) & 280 (20) & 0.5 (0.1) & -100 (10) & 340 (20) \\
   \feii & 17.94 &  --- & 0.3 (0.2) & -70 (10) & 860 (189) &  --- & --- & --- \\
   \siii & 18.71 & 23.3 & 3.3 (0.7) & -340 (10) & 900 (91) & 0.2 (0.0) & -100 (10) & 330 (20) \\
   \oiv & 25.89 & 54.9 & 12.2 (0.8) & -680 (10) & 580 (20) &  --- & --- & --- \\
 \enddata
 \tablecomments{Uncertainties on these fits, listed in parentheses next to the values, are calculated by \qtdfit\ using the residuals from the line fit for the flux and from the fit covariance matrix for the sigma. A minimum uncertainty of 10 \kms\ for $v_{50}$ and 20 \kms\ for $w_{80}$ has been imposed based on instrumental limits. We use these uncertainties for all lines fit with \qtdfit.}
\end{deluxetable*}
% nii 14.5, naiii 47.3
\begin{figure}
    \centering
    \includegraphics[width=\columnwidth]{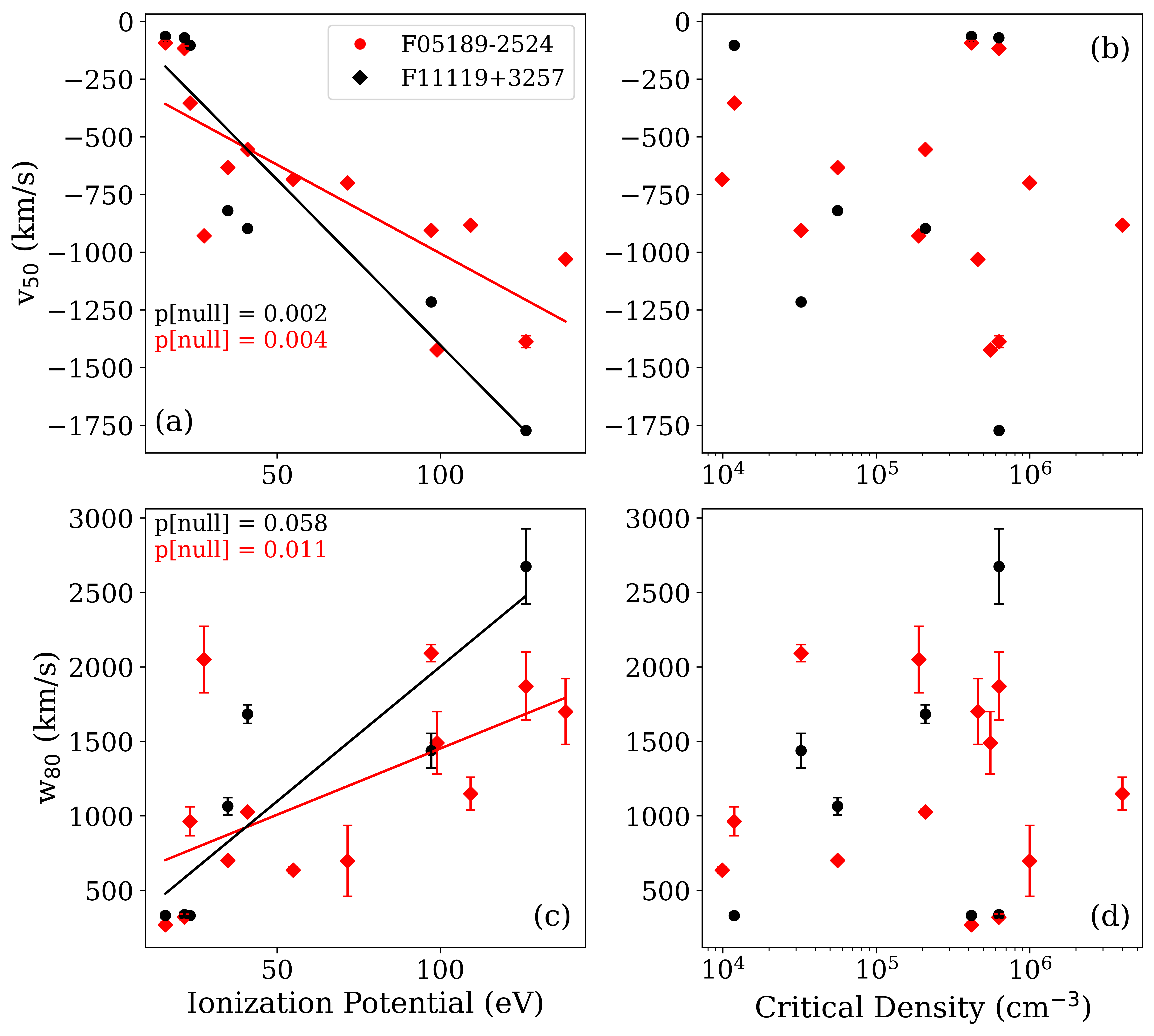}
    \caption{Left column: plot of $v_{50}$ (top left) and $w_{80}$ (bottom left) vs ionization potential. Right column: plot of $v_{50}$ (top right) and $w_{80}$ (bottom right) vs critical density. Data points represent all ionized gas emission lines fit in F11119+3257 (black) and F05189-2524 (red). The data for all plots are fit with a linear regression and any fits with a p[null] $< 0.1$ are displayed on the plot and the p[null] for significant fits is listed with the corresponding color for the correct source}
    \label{fig:IPvsv50}
\end{figure}

The positive correlation between $|v_{50}|$ and IP is best illustrated by the sequence of strong MIR neon emission lines (\neii\ 12.81 \um, \neiii\ 15.56 \um, \nev\ 14.32 \um, and \nevi\ 7.65 \um). We show the kinematic fits to four neon emission lines for both sources in Figure \ref{fig:NucNe}. The emission from both sources is remarkably similar. The lowest IP line \neii\ primarily shows narrow line emission at near systemic velocities with a possible weak outflow $v_{50}$ $\sim$ -1000 \kms\ potentially contaminated by emission from a PAH feature at 12.7 \um. The central peak of the remaining three lines is increasingly blueshifted with respect to the IP of the lines (up to $v_{50}$ $\sim$ -1500 \kms\ in \nevi), and the broad, blueshifted wing increases in strength and blueshift along the same sequence (up to $v_{90}$ $\sim$ -4000 \kms\ in \nevi). Evidence of systemic emission is still seen in \neiii\ but not in \nev\ or \nevi. These same trends are seen in other elemental series in both sources (e.g., argon, sulfur, and sodium) see Table \ref{tab:nuc_res} for more details. 

A few small differences are present between the emission of both sources. In all lines, the emission from F11119+3257 is generally slightly more blueshifted and broader, consistent with Figure \ref{fig:IPvsv50}. In \nev\ emission from F05189-2524, a distinct redshifted component is visible centered near 750 \kms; no such component is present in any of the F11119+3257 lines. The continuum noise in F05189-2524 is weaker than in F11119+3257, causing individual line components to be slightly more distinct in F05189-2524. 

\begin{figure}
    \centering
    \includegraphics[width=.7\columnwidth]{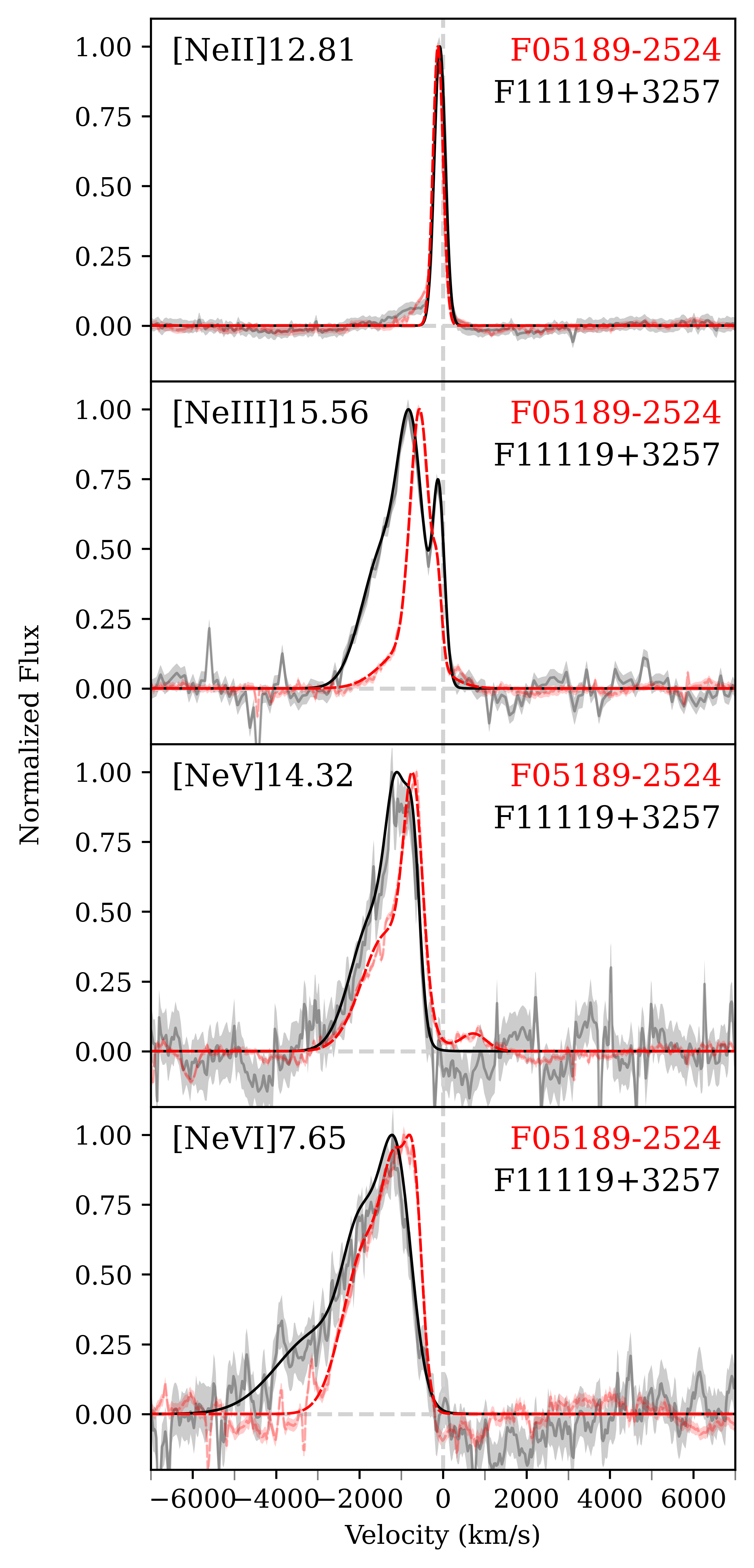}
    \caption{Fits to the MIR neon emission lines extracted from the nuclear, unresolved emission of F11119+3257 (red dashed) and F05189-2524 (black). The data are shown in light red dashed for F11119+3257 and gray for F05189-2524 with the 1-$\sigma$ error range filled in with the corresponding color for each source. The kinematics of each line are fit with up to three independent Gaussians. There is weak PAH emission at 12.7 \um\, just to the blue side of \neii, which is difficult to decouple from \neii\ emission.
    %Line emission categorized as outflow ($|v|\ge300$ \kms) is shaded in gray. 
    }
    \label{fig:NucNe}
\end{figure}

The molecular gas fits (H$_2$ $0 - 0$ S(1) $-$ S(8)) for both sources are kinematically very similar. They both show no clear signs of outflow and exhibit a slight blueshift, with average median velocities (for all lines) of $v_{50}^{F05189}$ $\sim$ -110 \kms and $v_{50}^{F11119}$ $\sim$ -60 \kms and are narrower than the ionized lines with average velocity widths of $w_{80}^{F05189}$ $\sim$ 250 \kms and $w_{80}^{F11119}$ $\sim$ 320 \kms. 

\subsection{Extranuclear Emission}
\label{subsec:res_extranuc}
 
\subsubsection{Low Velocity}
\label{subsec:res_molecular}

Both sources show evidence of extended emission of rotational H$_2$ and low IP warm ionized gas lines at low, near-systemic, velocities. We run a \qtdfit\ analysis of the rotational H$_2$ lines from S(1) to S(7) for both sources on the non-unified cubes. In F05189-2524, we additionally detect and fit the S(8) transition, but any higher transitions are outside of the MRS observed wavelengths. The higher redshift of F11119+3257 brings the S(9), S(10), and S(11) transitions into view, but they are too weak relative to the continuum noise to detect. For both sources, we fit the \arii\ 6.99 \um\ and \neii\ 12.81 \um\ ionized lines due to their strength, evidence for low-velocity extended emission, and because they lie in the same sub-bands as the S(5) and S(2) H$_2$ lines, making them good references for warm ionized gas emission. For these fits, we kinematically tie the velocity and velocity dispersion of the ionized gas emission to the molecular emission in order to reduce noise in the ionized gas fits as the lines trace kinematically similar emission but the molecular lines are stronger in the extended emission. We run the same \qtdfit\ analysis on the unified cubes, which give consistent results with those described below. The results of the fitting of two of the strongest and best resolved molecular lines, S(3) and S(5), are shown in Figure \ref{fig:H2} for both sources. We show maps of the quasar-subtracted flux, $v_{50}$, and $w_{80}$. 

\begin{figure*}
    \centering
    \includegraphics[width=\textwidth]{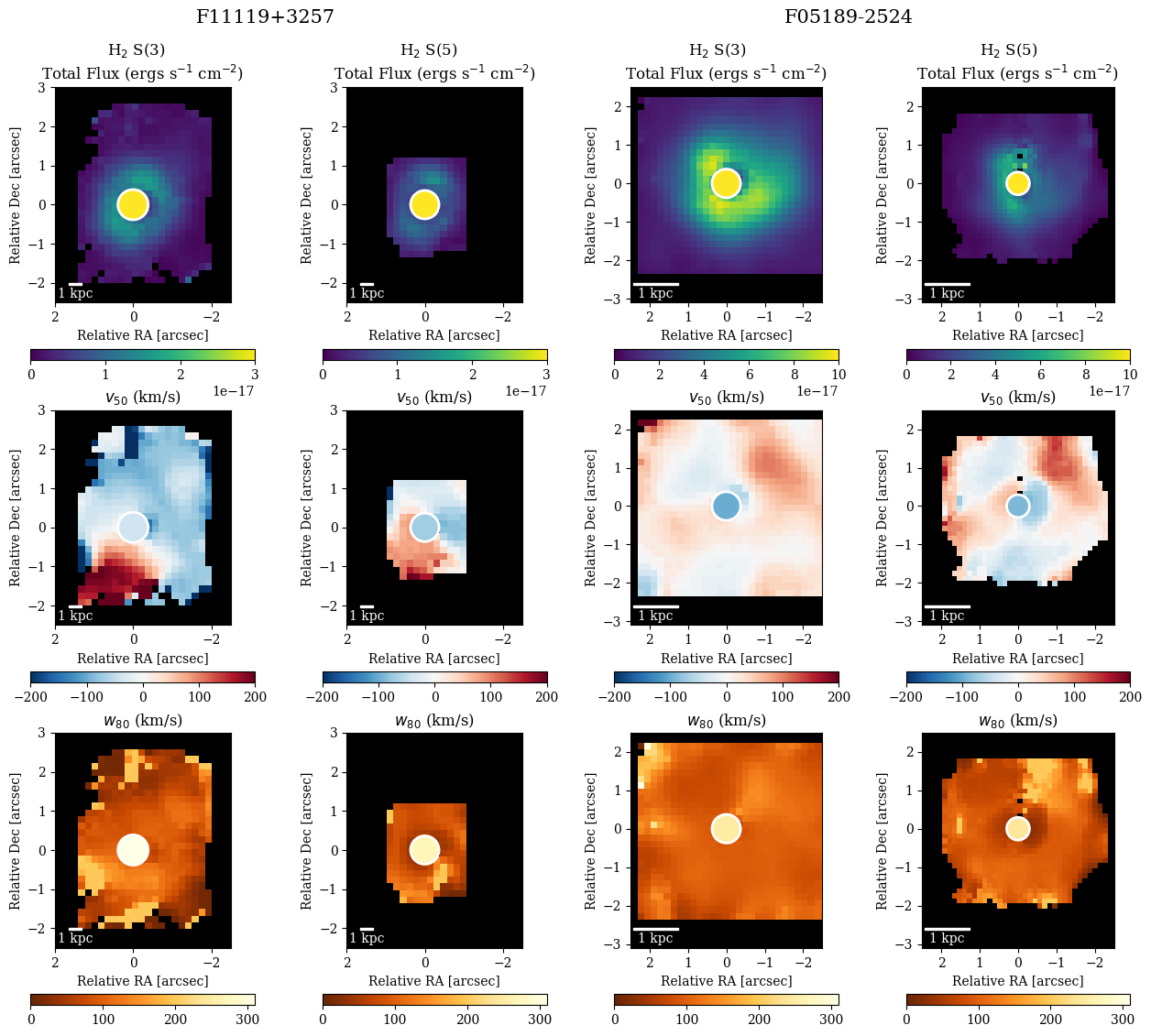}
    \caption{Flux and kinematic maps of two molecular lines H$_2$ $0 - 0$ S(3) and S(5) from the MIRI/MRS data cubes of both sources. The two left columns correspond to the emission from F11119+3257, and the two right columns correspond to the emission from F05189-2524. Each column shows maps of the emission line flux after processing with \qtdfit\ (quasar and stellar continuum emission removed), 50-percentile (median) velocities, $v_{50}$, and 80-percentile line widths, $w_{80}$, from top to bottom. Fits with a peak flux $\le1\sigma$ have been removed from the result. The white-outlined dot marks the quasar center with a size of the PSF FWHM. The color of the dot represents the corresponding parameter value fit in the nuclear spectrum.}
    \label{fig:H2}
\end{figure*}

For F11119+3257, the $v_{50}$ velocity field for both molecular lines shows a low-velocity gradient, $v_{50, red}\: - \:v_{50, blue} \sim 200$ \kms, from northwest (blueshifted) to the southeast (redshifted). This is likely attributed to disk rotation, confirmed by ALMA CO (1-0) measurements \citep{Vei2017}. The extent of this gradient and the excess emission are greater in S(3) than S(5), extending to $\sim$ 5 kpc from the nucleus at its greatest extent. Both lines show a slight decrement of flux in the inner kpc, which we attribute to an artifact of the PSF subtraction. Following methods used in \citet{Liu2020} derived from \citet{Vei2020}, we used velocity measurements from the best spatially and spectrally resolved line S(5) to measure a conservative lower limit on the circular and escape velocities of F11119+3257. We measure a $v_{\mathrm{circ}}=\sqrt{v_{\mathrm{H}_{2}}^2 + 2\sigma_{\mathrm{H}_{2}}^2}=130$ \kms\ and a $v_{\mathrm{esc}} \simeq 3v_{\mathrm{circ}}=390$ \kms\ by taking observed maximum values in the inner rotating disk. These methods give a conservative lower limit through assumptions of a density profile of a truncated single isothermal sphere, no inclination, and a constant density between the quiescent and outflowing gas. 

For F05189-2524, the larger spatial scale causes the excess emission to push the boundaries of the MRS FOV, extending up to $\sim$ 3 kpc in S(1), the line in the sub-band with the largest FOV. S(5) shows clear evidence of a low-velocity gradient, $v_{50, red}\: - \:v_{50, blue} \sim 100$ \kms, from the west (blueshifted) to the east (redshifted) in the inner kpc of the galaxy. This gradient is marginally seen in S(3), but the lower spatial resolution makes it difficult to confirm, as inner spaxels are more confounded by PSF removal. This gradient is confirmed by other high spatial resolution molecular lines \citep{Lut2020, Lam2022} and is consistent with stellar velocities \citep{Rup2017}. A spectrally broad, redshifted clump is seen $\sim$ 1 kpc to the northwest in both lines with a $v_{50}$ $\sim$ 100 \kms\ and a  $w_{80}$ $\sim$ 500 \kms. The rest of the field shows patchy blue and redshifted emission with $|v_{50}|$ $\le$ 100 \kms. Similar to F11119+3257, F05189-2524 shows a slight decrement in the flux of molecular lines in the region closest to the quasar which we attribute to a PSF subtraction artifact. 

Again using S(5) and the same methods as in F11119+3257, we calculate $v_{\mathrm{circ}}=110$ \kms\ and a $v_{\mathrm{esc}}=330$ \kms. We improve these estimates by using the absolute H-band magnitude of the host galaxy from \citep{Vei2006} to estimate the stellar luminosity of F05189-2524. Then, assuming a Milky Way-like structure, we scale up Milky Way rotational velocities by the square of the galaxy luminosity ratio and estimate $v_{\mathrm{circ}}=120$ \kms\ and $v_{\mathrm{esc}}\simeq360$ \kms\ for F05189-2524.

For both sources, the fit maps from molecular lines not shown in Figure \ref{fig:H2} show slight variations in spatial extent and velocities compared to the S(3) and S(5) lines, but the general conclusions are consistent with the results described above. 

To explore the decrement of molecular gas in the nuclear region of both sources, we compare the strength of spectrally adjacent low-velocity molecular and ionized gas lines in the same sub-bands. Figure \ref{fig:H2_ratio} shows the flux ratio of H$_2$ S(5) to \arii\ 6.99 \um\ and H$_2$ S(2) \neii\ 12.81 \um\ in log scale. 
%White pixels show regions where the molecular gas line was fit, but the ionized gas line was not. Black pixels show regions where neither line was fit. 
Both sources show an increase in the strength of molecular lines relative to the ionized emission lines as the distance from the central quasar increases. The molecular gas lines are 100 times weaker in the nuclear region than they are in the circumnuclear region relative to the ionized lines. This is strong evidence for AGN suppression of molecular gas and thus star formation in the nuclear region. 

\begin{figure}
    \centering
    \includegraphics[width=\columnwidth]{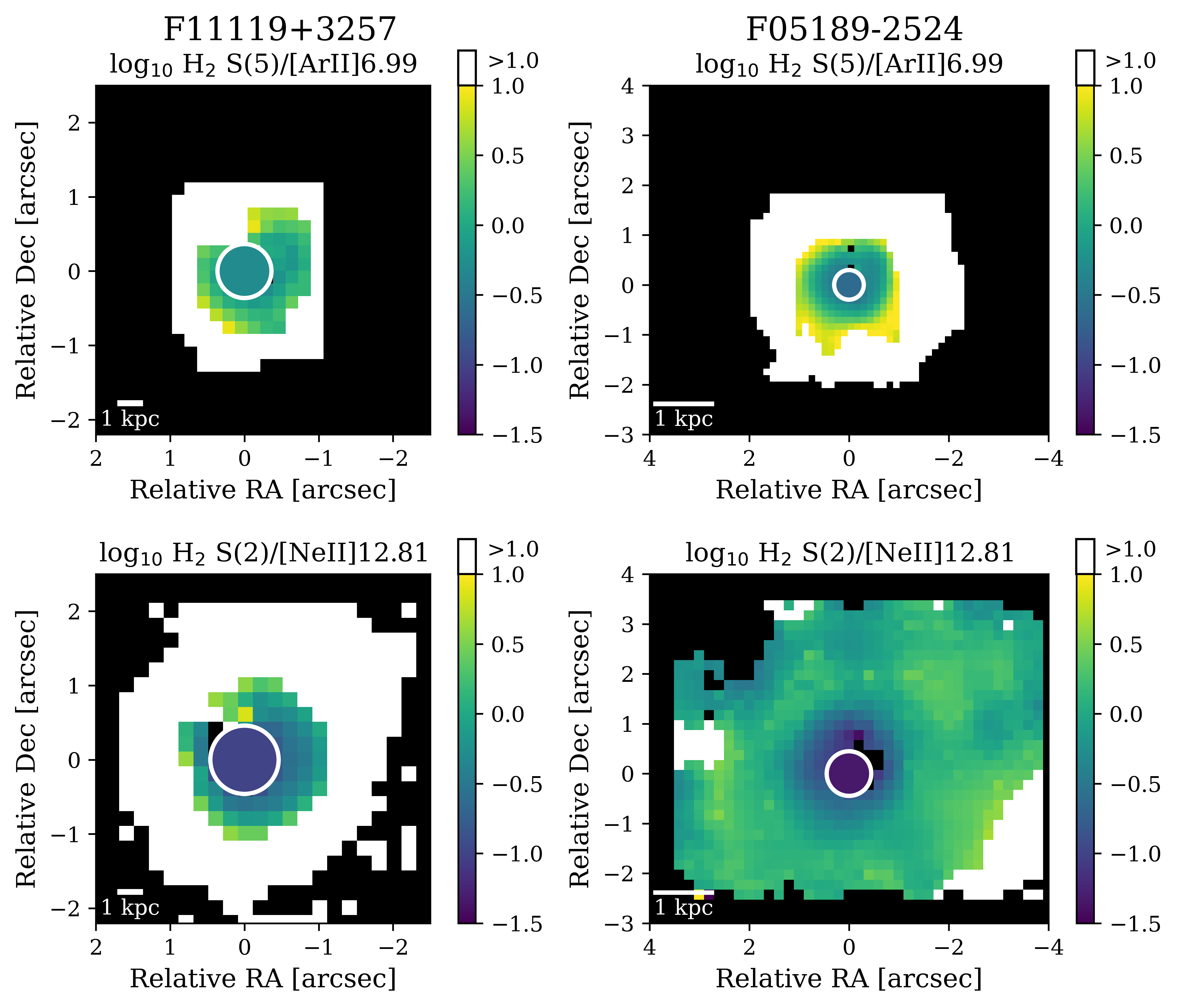}
    \caption{Emission line flux (quasar and stellar continuum emission removed) ratio of H$_2$ S(5)/\arii\ (top row) and H$_2$ S(2)/\neii\ 12.81 (bottom row) for F11119+3257 (left column) and F05189-2524 (right column). White pixels show regions with a ratio greater than 1, where molecular lines were fit, but ionized lines were not. Black pixels show regions where neither molecular nor ionized lines were fit. The white-outlined dot marks the quasar center with a size of the PSF FWHM. The color of the dot represents the corresponding parameter value fit in the nuclear spectrum.}
    \label{fig:H2_ratio}
\end{figure}

\subsubsection{High Velocity}
\label{subsec:res_ionized}

While the nuclear and low-velocity extended emission of both sources is remarkably similar, the high-velocity extended emission is where they begin to differ in our data. F05189-2524 shows clear evidence of high-velocity extended emission of high IP emission lines, while F11119+3237 shows no such evidence. The strongest evidence for this emission is seen in neon emission lines. We see no high-velocity extended emission in any of the other MIR fine structure lines.  

For F05189-2524 we run full cube \qtdfit\ analysis on \nevi\ 7.65 \um, \nev\ 14.32 \um, and \neiii\ 15.56 \um. We detect no significant outflowing emission in the \neii\ 12.81 \um\ line likely caused by the low velocities and contaminating PAH emission in the blue emission wing. In all of the neon lines, there is a dramatic shift of $\sim$ 2000 \kms\ from blueshifted emission in the north to redshifted emission in the south. \qtdfit\ is unable to range over such large velocities freely and is prone to fitting noise or missing components entirely when both the redshifted and blueshifted components are fit at the same time. To allow for reliable fitting and to reduce noise in the results, we only fit blueshifted emission to the north of the quasar and redshifted emission to the south. We do not see any significant redshifted emission to the north or blueshifted emission to the south that is being ignored by these methods. Using test fits of spaxels in both regions, we set initial guesses on the velocity, width, and number of Gaussian components for all lines. We fit two components to \nevi\ and \neiii\ and only one component to \nev\ in our fits to the extranuclear emission. 

The flux and kinematic maps resulting from this fitting are shown in Figure \ref{fig:Ion_map}. When multiple components are fit to the outflow, as in the case of \nevi, we combine them and calculate the flux, $v_{50}$, and $w_{80}$ for the combined complex. Examples of the kinematic spectral fits in the redshifted and blueshifted regions for both lines and \neiii\ are shown in Figure \ref{fig:Ion_spax}. 

A high-velocity, $|v_{50}|$ $\sim$ 1000 \kms, biconical emission is seen in all three neon emission lines in Figure \ref{fig:Ion_map}. This is well beyond the expected rotational velocity of the host galaxy ($\sim$ 150 \kms), and thus we classify it as an outflow. The general direction and velocities of this outflow are consistent with the visible \nii\ 6585 \AA\ measurements \citep{Rup2017}. CO measurements seem to show the same directionality as our outflow \citep{Flu2019, Lut2020}, although recent measurements from \citet{Lam2022} are not as decisive. All resolved molecular outflow detections show a strong redshifted component but very weak blueshifted component contradictory with our results. Similar to the nuclear emission, the outflow has a higher velocity in higher ionization stages, with a flux-weighted mean $v_{50}$ ranging from 500 \kms\ in \neiii\ to 810 \kms\ in \nevi. The data also show that the spatial extent of the outflow decreases in higher states, as the flux-weighted mean outflow radius R$_\mathrm{out}$ decreases from 1060 pc in \neiii\ to 860 pc in \nevi. Consistent with observations of another Seyfert 2 galaxy NGC 7319 \citep{Per2022}, and with a spatially stratified outflow as discussed in Section \ref{subsec:res_nuclear}.

\begin{figure*}
    \centering
    \includegraphics[width=0.8\textwidth]{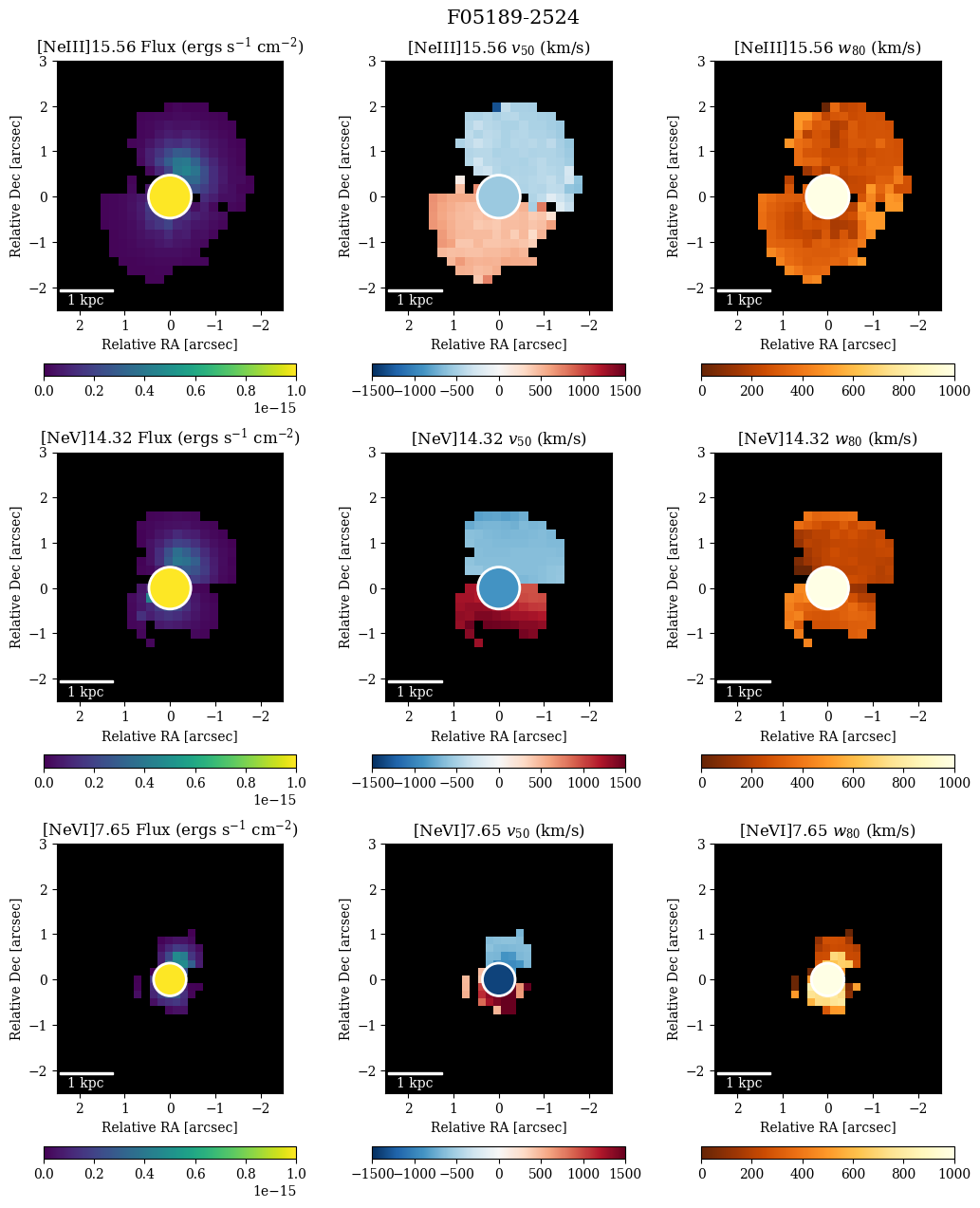}
    \caption{Flux and kinematic maps of three ionized lines \neiii\ 15.56 \um\ (top row), \nev\ 14.32 \um\ (middle row), and \nevi\ 7.65 \um\ (bottom row) from the MIRI/MRS data cube for F05189-2524. Each row shows maps of the emission line flux after processing with \qtdfit\ (quasar and continuum emission removed), 50-percentile (median) velocities, $v_{50}$, and 80-percentile line widths, $w_{80}$, from left to right. Fits with a peak flux $\le3\sigma$ have been removed from the result. The white-outlined dot marks the quasar center with a size of the PSF FWHM. The color of the dot represents the corresponding parameter value fit in the nuclear spectrum.}
    \label{fig:Ion_map}
\end{figure*}

The blueshifted outflow for all lines typically peaks at a slightly lower velocity ($\sim$ 100 \kms) than the unresolved emission and lacks an extended blueshifted tail. For the redshifted side \neiii\ and \nev\ are easily fit with a single Gaussian component, but \nevi\ shows two distinct clumps at $v_{50}$ $\sim$ 500 \kms and $v_{50}$ $\sim$ 1500 \kms. A low-velocity component, likely due to disk rotation, is also clearly visible, and fit, in \neiii\ but hard to quantify. In fitting of the redshifted outflow region, the quasar template often overfits the PSF emission. This can be seen in the bottom left and top left panels of Figure \ref{fig:Ion_spax} and only has a minor effect, slightly reducing derived masses.

\begin{figure*}
    \centering
    \includegraphics[width=\textwidth]{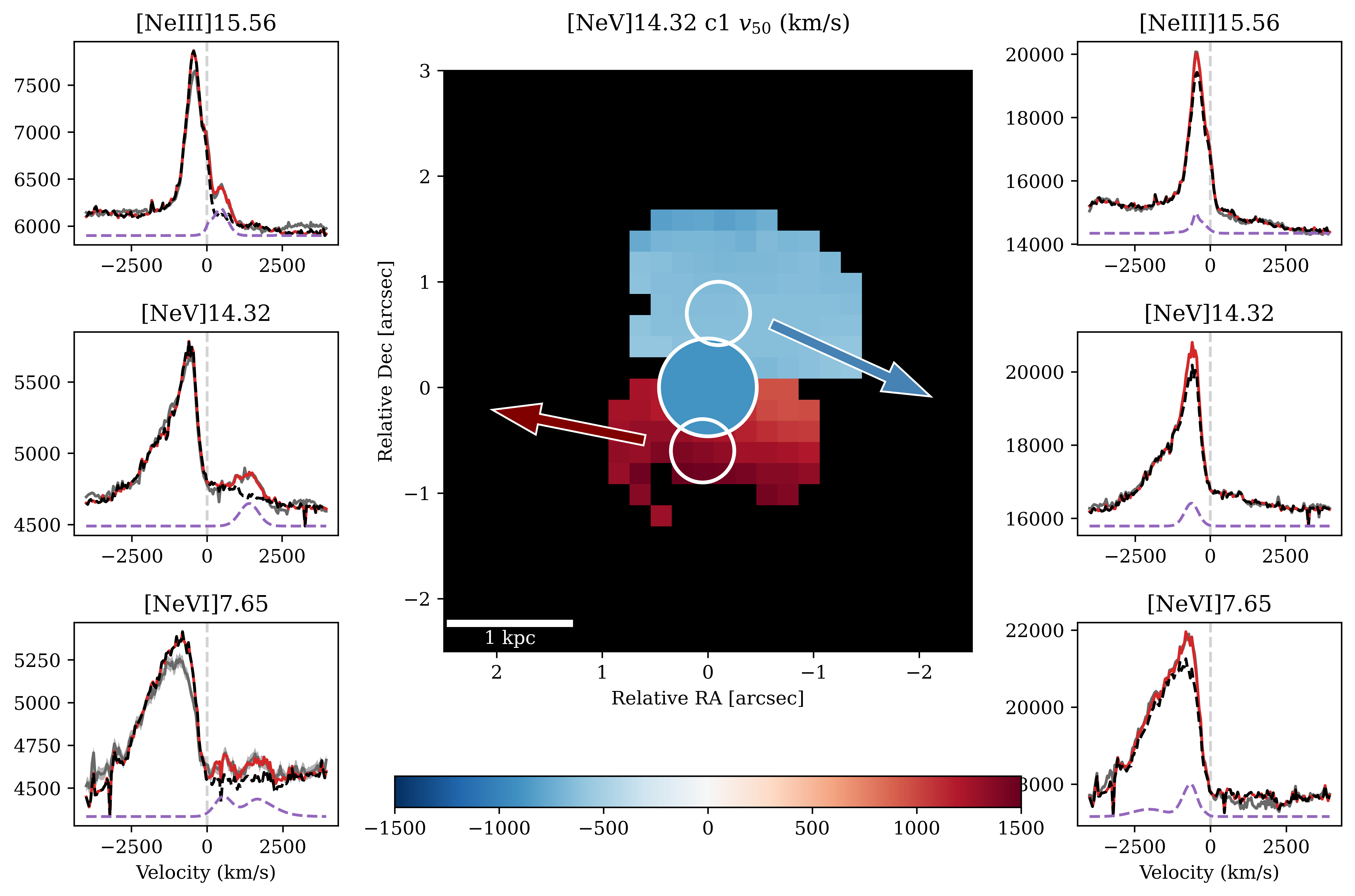}
    \caption{Examples of extracted spectra for three ionized lines, \neiii\ 15.56 \um, \nev\ 14.32 \um, and \nevi\ 7.65 \um, from the redshifted (left column) and blueshifted (right column) regions of the F05189-2524 spectral cube. The central map depicts the excess emission $v_{50}$ of \nev. Emission from the other two lines has been extracted for each marked circle from the corresponding spatial region of their respective cubes. Fits with a peak flux $\le3\sigma$ have been removed from the result. White unfilled circles mark the regions where spectra have been extracted from the redshifted (south of the quasar) and blueshifted (north of the quasar) emission regions. In the extracted spectra subplots, the red line is the sum of the scaled quasar template and the line fit, the latter of which is shown as the purple dashed line and the former of which is shown as the black dashed line. The data are the dark gray line (often hidden behind the red line) with the 1-$\sigma$ error range filled in with the lighter gray. The y-axis units are 10$^{-16}$\ergscm$\mu$m$^{-1}$.}
    \label{fig:Ion_spax}
\end{figure*}

For F11119+3257, we attribute the lack of detection of a resolved outflow to the greater distance to the source. The brightest portion of the \neiii\ outflow in F05189-2524 is $\le 1$ kpc from the quasar, which is the effective size of the PSF FWHM at that wavelength for F11119+3257, making the outflow difficult to resolve. This, in combination with the weaker \neiii\ emission in F11119+3257, makes resolved kpc-scale outflows unlikely to be observed with MIRI MRS. Still, given the similarity of the two sources and the strong unresolved warm ionized outflow, and errors involved, we cannot rule out the presence of a kpc-scale outflow in the source. 

\section{Discussion}
\label{sec:discussion}

\subsection{Dust}
\label{subsec:dis_dust}

The ratios between the different PAH features may be a good indicator of the PAH grain size, PAH ionization level, and incident radiation field on those PAHs \citep{Rig2021, Mara2022, Dra2021}. In Figure \ref{fig:pahratio}, we plot measured PAH fluxes from Table \ref{tab:PAH} against PAH model grids from \citet{Rig2021} and a sample of Seyfert galaxies from \citet{Gar2022a} and Seyfert and star forming galaxies from \citet{Rig2024} and \citet{Gar2022b}. For both sources, we find that F05189-2524 and F11119+3257 show evidence of a smaller ratio of 11.3 to 7.7 PAH than the Seyfert galaxies. This could either be explained by these sources having a more ionized PAH species or a PAH species under a stronger incident radiation field than the Seyfert galaxies. Additionally, we see that stronger ratios of 11.3 PAH and 6.2 PAH to the 7.7 PAH feature in F11119+3257, compared to those ratios in F05189-2524, are indicative of a smaller and less ionized PAH species. However, large errors in the measurements make conclusions indefinite. 

\begin{figure}
    \centering
    \includegraphics[width=\columnwidth]{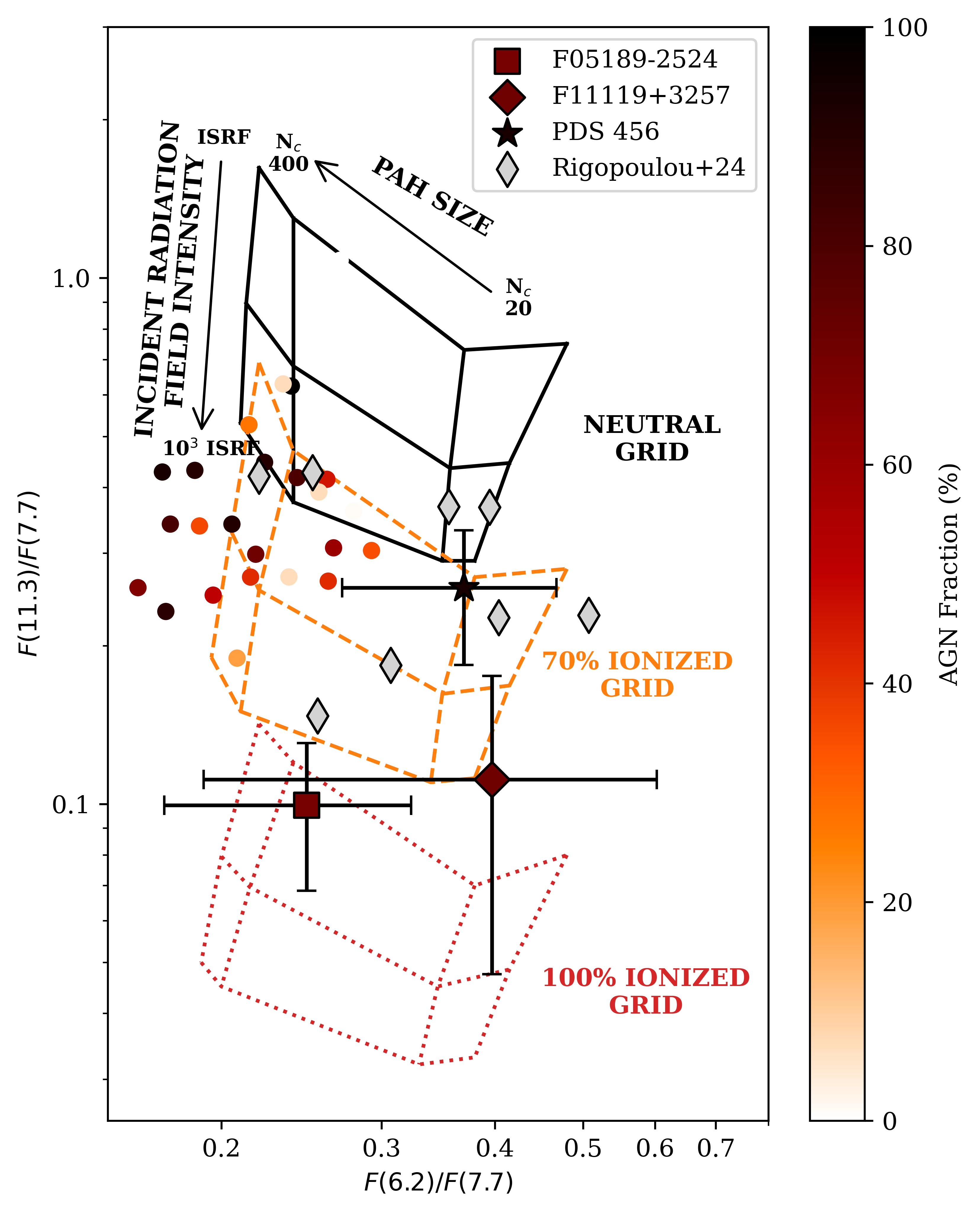}
    \caption{PAH ratio diagnostic diagram. Flux ratios of the PAH 6.2/7.7 \um\ and PAH 11.3/7.7 \um\ from Table \ref{tab:PAH}. The grids are estimated from \cite{Rig2021}. They correspond to mixtures of PAH species ionized to various levels (0\%, 70\%, and 100 \% ionized PAH molecules). The individual grids range from softer (1 $\times$ Interstellar Radiation Field (ISRF); top of the grid) to harder ($10^3$ $\times$ ISRF; bottom of the grid) incident radiation fields and from smaller ($N_c$ = 20, the number of carbon atoms; right side of the grid) to larger ($N_c$ = 400; left side of the grid) PAHs. The colored circles are PAH ratios for Seyfert galaxies colored by their MIR AGN fraction as defined in \cite{Gar2022a}. Values for F11119+3257, F05189-2524, and PDS 456 \citep{See2024} are plotted with error bars, and the color reflects their measured AGN contribution \citep{Vei2009, See2024}. \jwst\ observations of Seyfert and star-forming galaxies from \citet{Rig2024} are represented by the thin gray diamonds with the inclusion of NGC 6552 from \citet{Gar2022b}.}
    \label{fig:pahratio}
\end{figure}

The MIR PAH features are ubiquitous in galactic and extragalactic sources. These features are strongly correlated with SFR in normal non-active galaxies \citep{Pee2004, Shi2016, Li2020}. However, AGN suppression of PAH features can affect the SFR-L$_{\mathrm{PAH}}$ relation. This is done through the destruction of PAH molecules \citep{Voi1992, Sei2004}. 
%or on very small scales $\sim10-500$ pc \citep{Jen2017} possible excitation of PAH features by AGN radiation. 
The 11.3 \um\ PAH feature is arguably the least AGN-suppressed PAH feature in our observations and can fairly reliably indicate SFR for galaxies with AGN \citep{Dia2012, Shi2016}. 

Thus we use Equation 2 from \cite{Dia2012} to derive SFR = 6 $\pm$ 2 M$_{\odot}$ yr$^{-1}$ for F11119+3257 and SFR = 4 $\pm$ 2 M$_{\odot}$ yr$^{-1}$ for F05189-2524. We find these results on the lower end of other SFR-L$_{\mathrm{PAH}}$ relations using the other PAH features (6.2 \um\ and 7.7 \um) and calibration samples \citep{Far2007, Pop2008, Lut2008, Tre2010, Shi2016}; which range from 4 -- 114 M$_{\odot}$ yr$^{-1}$ for F11119+3257 and 4 -- 81 M$_{\odot}$ yr$^{-1}$ for F05189-2524. We attribute this large range to variations in PAH--SFR relation calibration samples and methods, see \citealt{Shi2016} for a more detailed discussion. For F11119+3257, the high-end results from this analysis are consistent with the SFR estimated from FIR dust emission 190 $\pm$ 90 M$_{\odot}$ yr$^{-1}$ \citep{Pan2019}. 
%, but not with the SFR estimated from PAH features 600 $\pm$ 300 M$_{\odot}$ yr$^{-1}$. 

\subsection{Molecular Gas}
\label{subsec:dis_molecular}

To each spaxel in the unified cube fits, using emission probabilities for the H$_2$ $v = 0 - 0$ transitions from \cite{Rou2019} and Equation 1 of \cite{You2018} to calculate $N(\nu_u, J_u)$, we construct an excitation diagram for all of the H$_2$ $0-0$ lines present in the spaxel. We then implement two methods to determine the temperature and column density of the molecular gas. We provide an example fit using both of these methods in Figure \ref{fig:H2_nuc_excitation} using the data from the nuclear emission.  

The first method uses Equation 2 from \cite{You2018},
\begin{equation}
\label{eqn:h2temp}
    log_{10} \frac{N(v_u, J_u)}{g(J_u)}=-\frac{1}{T\cdot ln(10)} \frac{E(v_u, J_u)}{k_B} + log_{10} N(0,0), 
\end{equation}
to construct a two-component linear fit of the data. The first component fits lines sensitive to cooler temperatures (S(1), S(2), S(3); T$_{cool}$) and the second component fits only the lines sensitive to hotter temperatures (S(4), S(5), S(6), S(7), S(8); T$_{warm}$). We use the column densities of the mass-dominant, cooler molecular gas to calculate the per-spaxel H$_2$ gas mass. This broken linear fit can be seen as the blue (cold) and red (warm) lines in Figure \ref{fig:H2_nuc_excitation}. 

The second method uses an adapted Python implementation from \citet{Jon2024} of the H$_2$ excitation model of \citet{Tog2016} to fit the molecular data. The \citet{Tog2016} model assumes a power-law distribution for rotational molecular emission lines and fits the lower bounds on the temperature distribution $T_l$ and the power law slope. These, in combination with an assumed upper bound on the temperature distribution ($T_u$ = 2000 K; the model is insensitive to any values larger than 2000 K), allow for the total column density and mass of H$_2$ to be fit. This power law fit can be seen as the gray solid line in Figure \ref{fig:H2_nuc_excitation}. 

We note that the \citet{Tog2016} method can better extrapolate the S(0) flux, which is unobservable with MIRI and an indicator of cooler, more abundant gas. This causes the method to provide consistently greater column densities and lower temperature estimates than the broken linear method, but we are unable to consistently quantify this difference across varying spaxels in order to apply a correction.

In all spaxels and the nuclear fit, we assume an ortho/para ratio of 3 for the statistical weights of the molecular transitions. Smaller ortho/para ratios have been observed in some cases \citep[e.g.][]{Hab2005}, but a value of 3 is a conservative upper limit useful for comparisons, and we generally find that lower values cause a discontinuity between the ortho and para transitions.

In the nuclear region, we see a significant drop in the S(3) line relative to other nearby lines. We attribute this to extinction from the silicate absorption feature from $\tau_{9.7}$. We correct this by fitting the \citet{Tog2016} method with all lines excluding S(3), correcting S(3) to the \citet{Tog2016} fit, and refitting the broken linear method. The uncorrected S(3) line is shown in light gray in Figure \ref{fig:H2_nuc_excitation} while the corrected version is shown in black. Outside of the nucleus, we see no significant signs of extinction and lack the data quality to consistently determine and correct for it. Thus, we do not correct the H$_2$ S(3) line for extinction in our extranuclear fits. 

\begin{figure}
    \centering
    \includegraphics[width=\columnwidth]{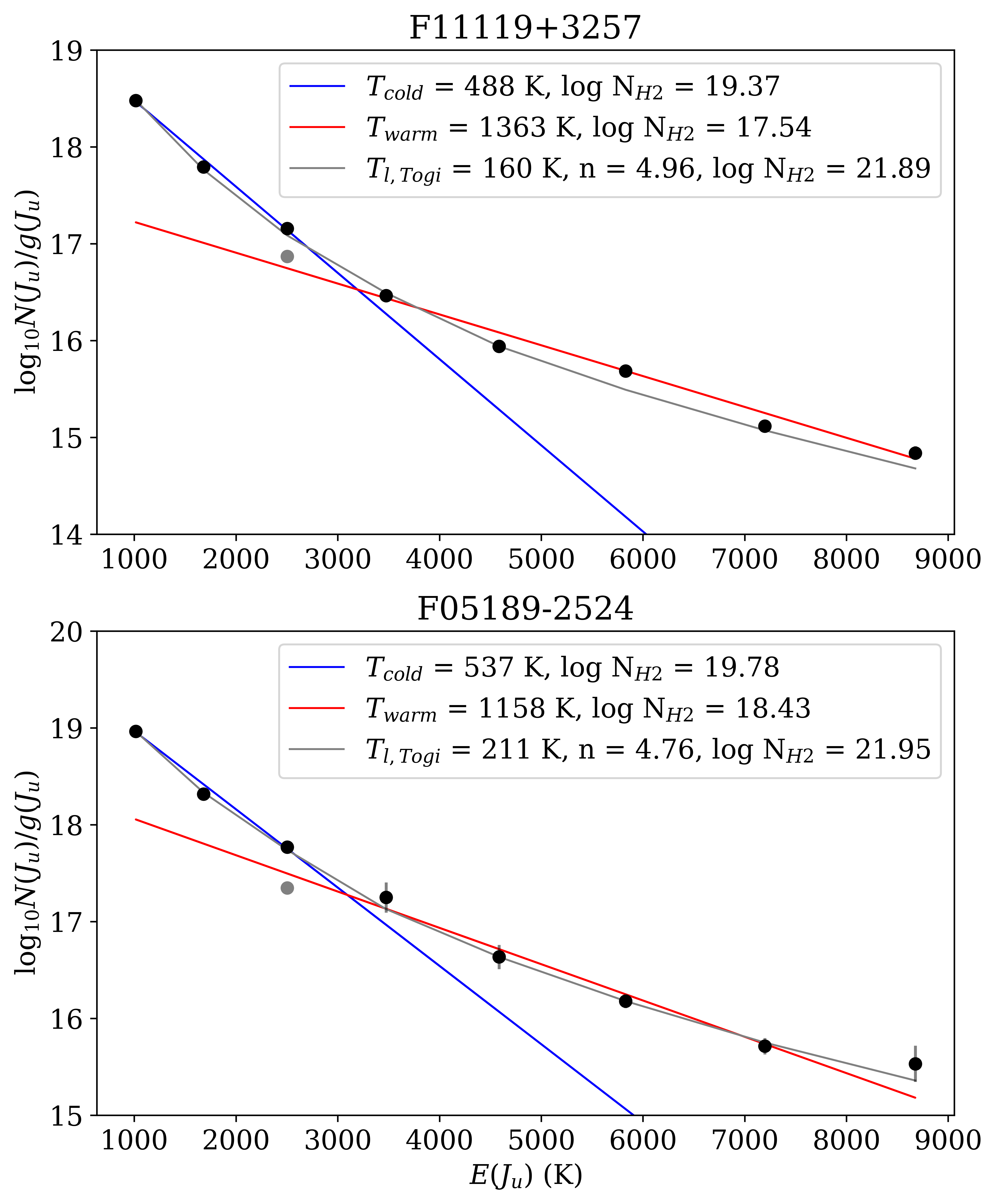}
    \caption{Excitation diagram for the rotational transitions of H$_2$ for F11119+3257 (top) and F05189-2524 (bottom) extracted from the nuclear spectrum. Values of $g(J_u)$, $E(J_u)$, and $A$ were gathered from \cite{Rou2019} in order to convert from line intensities to $N(J_u)$. We fit the data points with two methods: one using two-component (warm and cold) linear regression and two a power-law distribution based on the model in \citet{Tog2016}. The uncorrected for extinction emission for H$_2$ is represented by the gray point, and the corrected emission by the black point. }
    \label{fig:H2_nuc_excitation}
\end{figure}

From our full cube fitting, we find that the \citet{Tog2016} power-law method is very susceptible to fitting errors with this data. This occurs largely in the H$_2$ S(1) line caused by nearby fringing or residual noise and can lead to unphysically high column densities and low temperatures. In addition, the lower signal S(4) $-$ S(7) lines often confound the fitting. We find that the simple linear fit to colder warm molecular gas is more robust in response to fitting errors and does not require the weaker S(4) $-$ S(7) lines.  

This leads us to use the linear fit on the lines sensitive to cooler temperatures (S(1), S(2), S(3)) in our full temperature and column density maps shown in Figure \ref{fig:H2_Tne}. To make these maps, we only fit spaxels where all three lines are present above 1-$\sigma$ as depicted by the white region in the right column of Figure \ref{fig:H2_Tne}. We display the nuclear fitting results as a circle with the size PSF FWHM and colored based on the nuclear temperature and column density. Both sources show higher column density and temperature in the nuclear region as compared to the extended emission. In F11119+3257, we see no significant spatial trends in temperature or column density. In F05189-2524, we note that the broad, slightly redshifted clump to the northwest shown in Figure \ref{fig:H2} is spatially correlated with a slight decrease in column density with respect to nearby regions. 

Using the column density results from both the unresolved and resolved cases, we can extract a total mass in warm molecular gas by multiplying by the total area uniquely subtended by the unresolved and resolved emission and the mass of a $H_2$ molecule. For F05189-2524, we derive a mass of $M_{H2}^{warm} =4.5\pm1\times10^8$ $M_{\odot}$ with 1.3\% of the mass unresolved. This makes up $\sim 3\%$ of the cold molecular mass traced by CO line emission \citep{Flu2019, Lut2020, Lam2022}. For F11119+3257, we derive a mass of $M_{H2}^{warm}=3.0\pm1\times10^8$ $M_{\odot}$ with 0.2\% of the mass unresolved. This makes up $\sim 3\%$ of the cold molecular mass traced by CO line emission \citep{Vei2017}. For both sources, we find our results in line with previous measurements of $M_{H2}^{warm}$ \citep{Vei2009, Per2014} and the comparison between warm and cold molecular gas \citep{Rig2002, Per2014}. 

% F11119 This paper Mh2 warm = 1.7e8 (rotational H2 lines)
% F11119 Veilleux09 Mh2 warm= 1.86e9 (rotational H2 lines)
% F11119 Veilleux17 Mh2 cold = 1e10 (CO (1-0))

% F05189 This paper Mh2 warm = 1.6e8 (rotational H2 lines)
% F05189 Veilleux09 Mh2 warm = 1.07e8 (rotational H2 lines)
% F05189 Pereira-Santaella14 log NH2 warm = 19.5 (rotational H2 lines)
% F05189 This paper log NH2 warm = 19.8 (rotational H2 lines)
% F05189 Pereira-Santaella14 log NH2 cold = 21.6 (CO measurements)
% F05189 Fleutsch19 Mh2 cold = 1e9 (CO (3-2))
% F05189 Mh2 cold for disk not clear from Lutz 2020

\begin{figure*}
    \centering
    \includegraphics[width=\textwidth]{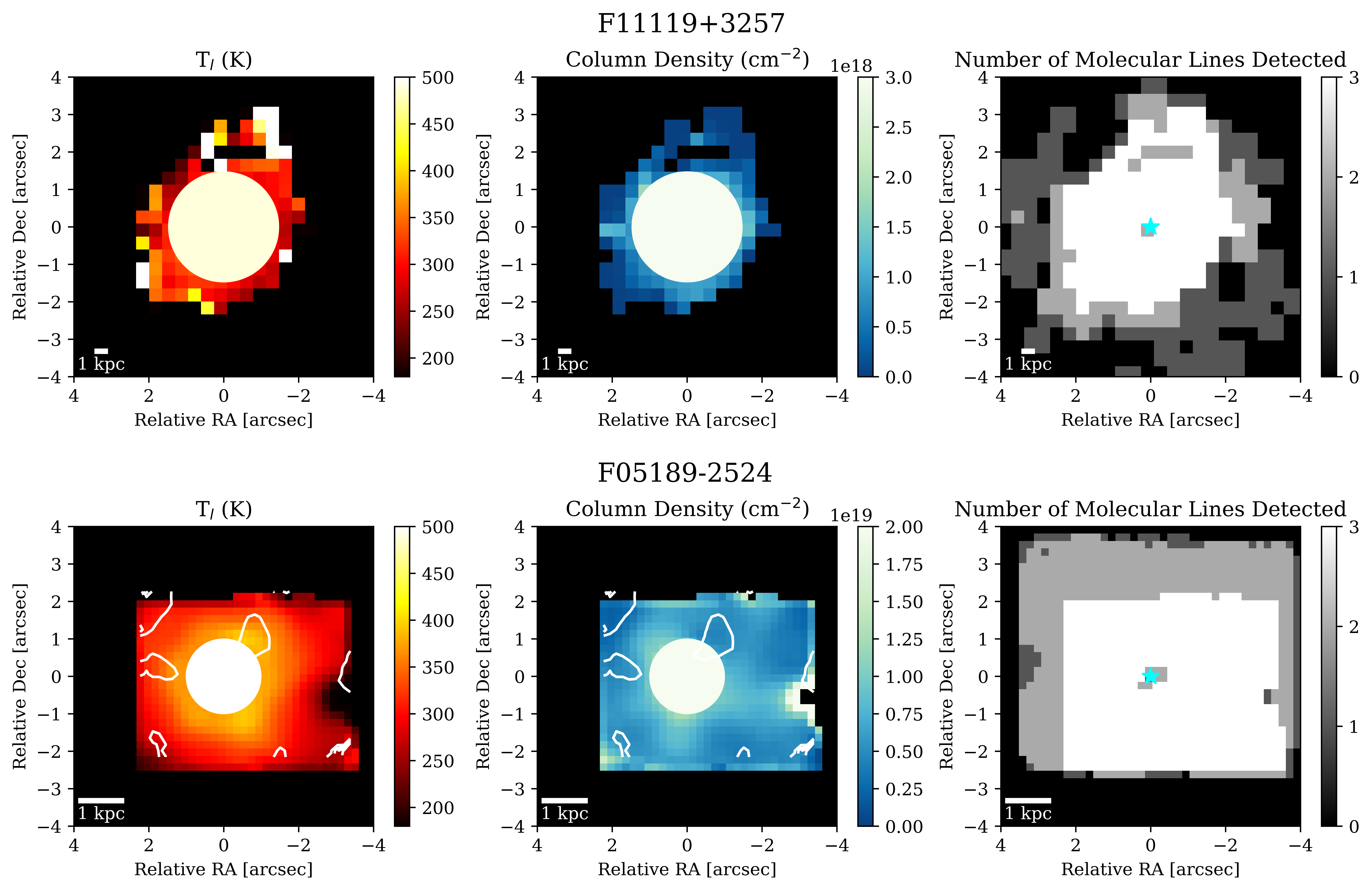}
    \caption{Maps of the temperature (K; left column), column density (cm$^{-2}$; center column), and number of molecular lines detected (right column) in F11119+3257 (top row) and F05189-2524 (bottom row). Temperatures and column densities are determined using a linear fit to the rotational H$_2$ lines sensitive to cooler temperatures (S(1), S(2), S(3)). Only spaxels where all three lines are present are fit. The central circles in the left and center columns mark the quasar center with a size of the PSF FWHM and color of the corresponding parameter value. The white contour in the temperature and column density maps of F05189-2524 traces the velocity width of the extended H$_2$ S(3) emission shown in Figure \ref{fig:H2}}
    \label{fig:H2_Tne}
\end{figure*}

\subsection{Ionized Gas}
\label{subsec:dis_ionized}

%%%%%%%%%%%%%%%%%%%%%%%%%%%%%%%%
We use the luminosities of the outflowing neon-emitting gas to estimate the mass of the warm ionized outflow from our data. For every spaxel, we take the fit of the extranuclear component and sum all of the flux with absolute velocities greater than 300 \kms, then convert the total flux of the outflow into a luminosity using the luminosity distance $D_L$. We choose $\pm300$ \kms\ as a cutoff because it is beyond the observed projected disk rotational velocities from rotational hydrogen lines ($v_{\mathrm{circ}} \sim 110$ \kms\ for F05189-2524 and 130 \kms for F11119+3257). For the unresolved outflow component, we apply the same method to the fit of the nuclear spectrum. Then subtract the flux from the extranuclear emission so that it is not double-counted. 

Next, the nuclear and extranuclear line luminosities are used to estimate the mass in warm ionized gas, following a method similar to that in \citet{See2024}, except we use gas in the Ne$^{1+}$, Ne$^{4+}$, and Ne$^{5+}$ states in addition to the Ne$^{2+}$ state. We determine the amount of ionized gas explained by the emission of each neon species, then sum them together for a total warm ionized gas mass. To do this we assume four things: 1) the outflowing neon emission is only concentrated in these four states; 2) $n_\mathrm{H}=10\times n_\mathrm{He}$; 3) a solar neon abundance of [Ne/H] = $-$3.91 \citep{Nic2017}; and 4) line emissivities calculated from PyNeb \citep{Lur2015} (assuming constant $T = 10^4$ K, which is appropriate for AGN photoionized gas, and constant $n_e = 10^3$ cm$^{-3}$ consistent with measurements in \citealt{Rup2017}). This results in
\begin{equation}
\label{eqn:massneii}
    M_{ionized}^{\mathrm{[NeII]}} = 6.09 \times 10^8 \frac{C\:L_{44}(\neiii) }{\langle n_{e,3} \rangle 10^{\mathrm{[Ne/H]}}} M_{\odot},
\end{equation}\begin{equation}
\label{eqn:massneiii}
    M_{ionized}^{\mathrm{[NeIII]}} = 2.96 \times 10^8 \frac{C\:L_{44}(\neiii) }{\langle n_{e,3} \rangle 10^{\mathrm{[Ne/H]}}} M_{\odot},
\end{equation}
\begin{equation}
\label{eqn:massnev}
    M_{ionized}^{\mathrm{[NeV]}} = 2.99 \times 10^7 \frac{C\:L_{44}(\nev) }{\langle n_{e,3} \rangle 10^{\mathrm{[Ne/H]}}} M_{\odot},
\end{equation}
\begin{equation}
\label{eqn:massnevi}
    M_{ionized}^{\mathrm{[NeVI]}} = 4.17 \times 10^7 \frac{C\:L_{44}(\nevi) }{\langle n_{e,3} \rangle 10^{\mathrm{[Ne/H]}}} M_{\odot},
\end{equation}
where $C \equiv \langle n_{e} \rangle^2/\langle n_{e}^2 \rangle $ is the electron density clumping factor, which we assume is of order unity, $L_{44}$ is the neon luminosity normalized to 10$^{44}$ \ergs, $\langle n_{e,3} \rangle$ is the average electron density normalized to 10$^3\:\mathrm{cm}^{-3}$. As mentioned earlier, the mass from this estimate assumes a constant temperature T $\sim10^4$ K,  but note that the temperature dependence is weak. It also assumes that the electron densities are below the critical densities\footnote{
$n_{\mathrm{crit}}$(\neii) = $6.3 \times 10^5$ cm$^{-3}$ \newline
$n_{\mathrm{crit}}$(\neiii) = $2.1 \times 10^5$ cm$^{-3}$ \newline
$n_{\mathrm{crit}}$(\nev) = $3.2 \times 10^4$ cm$^{-3}$ \newline
$n_{\mathrm{crit}}$(\nevi) = $6.3 \times 10^5$ cm$^{-3}$} so that collisional de-excitation is unimportant. Summing these individual components gives the final derived ionized gas mass 
\begin{equation}
\label{eqn:massoutflow}
    M_{ionized} =  M_{ionized}^{\mathrm{[NeII]}} + M_{ionized}^{\mathrm{[NeIII]}} + M_{ionized}^{\mathrm{[NeV]}} + M_{ionized}^{\mathrm{[NeVI]}}.
\end{equation}

For each spaxel, we then use these calculated masses, the pixel scale $\Delta r$, and the outflow velocity $v_{out}\cong v_{50} + \sigma \sim v_{84}$ to roughly estimate the mass outflow rate,  

\begin{equation}
\label{eqn:massoutflow}
    \dot M_\mathrm{out} = \frac{M_\mathrm{out} v_{\mathrm{out}}}{\Delta r},
\end{equation}
momentum outflow rate, 
\begin{equation}
\label{eqn:momentumoutflow}
    \dot P_\mathrm{out} = \dot M_\mathrm{out} v_{\mathrm{out}},
\end{equation}
and outflow power, 
\begin{equation}
\label{eqn:energyoutflow}
    \dot E_\mathrm{out} = \frac{1}{2} \dot M_\mathrm{out} v_{\mathrm{out}}^2.
\end{equation}

For the nuclear emission, the FWHM/2 of the PSF is used as an upper limit on the $\Delta r$, and the rest of the calculation is the same. Once these energetics are calculated for each spaxel, the results for each neon state and the total warm ionized gas are summed, calculated, and presented in Table \ref{tab:outlflow}. For F11119+3257, we calculate masses only for the unresolved nuclear emission, but in F05189-2524, we calculate masses for both the resolved and unresolved emission. In addition, Table \ref{tab:outlflow} includes the outflow radius $R_{\mathrm{out}}$, outflow velocity $v_{out}$, and compiled results from other papers displaying the outflow energetic calculations from other species of gas derived at a wide variety of distances for both sources. For resolved emission, the flux weighted mean for all spaxels is reported.

\begin{deluxetable*}{c c c c c c c c c c}
 \tablecaption{Outflow Parameters
 \label{tab:outlflow}}
 \tablehead{\colhead{Ref} & \colhead{Gas Phase} & \colhead{Tracer} & \colhead{Component} & \colhead{$M_{\mathrm{out}}$} & \colhead{$v_{\mathrm{out}}$} & \colhead{R$_\mathrm{out}$} & \colhead{$\dot M_{\mathrm{out}}$} & \colhead{$\dot P_{\mathrm{out}}$} & \colhead{$\dot E_{\mathrm{out}}$}\\
 \colhead{} & \colhead{} & \colhead{} & \colhead{} & \colhead{(10$^5\:M_{\odot}$)} & \colhead{(\kms)} & \colhead{(pc)} & \colhead{($M_\odot \mathrm{yr}^{-1}$)} & \colhead{($10^{33} dyn$)} & \colhead{($10^{41}$ \ergs)}\\
 \colhead{(1)} & \colhead{(2)} & \colhead{(3)} & \colhead{(4)} & \colhead{(5)} & \colhead{(6)} & \colhead{(7)} & \colhead{(8)} & \colhead{(9)} & \colhead{(10)}\\
 \colhead{} & \colhead{} & \colhead{} & \colhead{} & \colhead{F11119-2524} & \colhead{} & \colhead{} & \colhead{} & \colhead{} & \colhead{}}
 \startdata
    (1) & Warm Ionized & \neii\ 12.81 & Unresolved & 29.4$_{-1.6}^{+1.6}$ & 1300$_{-100}^{+100}$ & 960 & 4.0$_{-0.2}^{+0.2}$ & 31.8$_{-1.8}^{+1.8}$ & 20.2$_{-1.2}^{+1.2}$\\
    (1) & & \neiii\ 15.56 & Unresolved & 36.7$_{-2.5}^{+2.5}$ & 1700$_{-100}^{+100}$ & 1150 & 5.1$_{-0.4}^{+0.4}$ & 53.8$_{-4.3}^{+4.3}$ & 45.3$_{-4}^{+4}$\\
    (1) & & \nev\ 14.32 & Unresolved & 1.8$_{-0.3}^{+0.3}$ & 1900$_{-200}^{+200}$ & 1050 & 0.3$_{-0.06}^{+0.06}$ & 3.9$_{-0.7}^{+0.7}$ & 3.7$_{-0.7}^{+0.7}$\\
    (1) & & \nevi\ 7.65 & Unresolved & 6.0$_{-3.3}^{+3.3}$ & 3100$_{-400}^{+400}$ & 670 & 2.8$_{-1.8}^{+1.8}$ & 53.5$_{-36.1}^{+36.1}$ & 82.0$_{-63.6}^{+63.6}$\\
    (1) & & & Total & 73.88$_{-5.5}^{+5.5}$ & & & 12.1$_{-1.8}^{+1.8}$ & 143.1$_{-18.5}^{+18.5}$ & 151.3$_{-27.1}^{+27.1}$\\
    (2) & Hot Ionized & Fe K & Unresolved & --- & 75900 & 2.3 $\times$ $10^{-5}$ & 0.5-2 & 240-1000 & 9100-38000\\
    (3) & Neutral & Na I D & Unresolved & 0.05 & 1000 & 10 & 0.59 & 25 & 13\\
    (4) & Cold Molecular & OH & Unresolved & 2350$_{-1100}^{+2700}$ & 1000$_{-200}^{+200}$ & 300 & 800$_{-550}^{+1200}$ & 5500$_{-3750}^{+7050}$ & 2750$_{-2050}^{+4780}$\\
    (5) & Cold Molecular & CO (1-0) & Unresolved & 6000-14000 & 1000$_{-200}^{+200}$ & 4000-15000 & 80-200 & 750-1500 & 300-900\\
    \\
     & & & & F05189-2524 & & & & & \\
    \hline
    (1) & Warm Ionized & \neii\ 12.81 & Unresolved & 9.00$_{-0.24}^{+0.24}$ & 700$_{-60}^{+60}$ & 240 & 2.73$_{-0.27}^{+0.27}$ & 12.16$_{-1.69}^{+1.69}$ & 4.29$_{-0.9}^{+0.9}$\\
    (1) & & & Resolved & --- & --- & --- & --- & --- & ---\\
    (1) & & \neiii\ 15.56 & Unresolved & 14.19$_{-0.67}^{+0.67}$ & 950$_{-80}^{+80}$ & 270 & 5.11$_{-0.65}^{-0.65}$ & 30.46$_{-5.29}^{+5.29}$ & 14.42$_{-3.86}^{+3.86}$\\
    (1) & & & Resolved & 1.45$_{-0.34}^{+0.34}$ & 700$_{-360}^{+625}$ & 1060$_{-950}^{+1010}$ & 0.61$_{-0.14}^{+0.14}$ & 3.06$_{-0.71}^{+0.71}$ & 1.37$_{-0.36}^{+0.36}$\\
    (1) & & \nev\ 14.32 & Unresolved & 1.94$_{-0.24}^{+0.24}$ & 1600$_{-100}^{+199}$ & 250 & 1.25$_{-0.12}^{+0.12}$ & 12.50$_{-1.48}^{+1.48}$ & 9.96$_{-1.71}^{+1.71}$\\
    (1) & & & Resolved & 0.14$_{-0.03}^{+0.03}$ & 920$_{-850}^{+880}$ & 710$_{-620}^{1190}$ & 0.07$_{-0.02}^{+0.02}$ & 0.42$_{-0.10}^{+0.10}$ & 0.19$_{-0.05}^{+0.05}$\\
    (1) & & \nevi\ 7.65 & Unresolved & 2.25$_{-0.83}^{+0.83}$ & 2000$_{-200}^{+200}$ & 170 & 2.73$_{-0.99}^{+0.99}$ & 34.78$_{-13.35}^{+13.35}$ & 35.15$_{-15.03}^{+15.03}$\\
    (1) & & & Resolved & 0.13$_{-0.04}^{+0.04}$ & 1430$_{-1000}^{+1350}$ & 860$_{-410}^{+600}$ & 0.11$_{-0.03}^{+0.03}$ & 1.24$_{-0.32}^{+0.32}$ & 1.24$_{-0.31}^{+0.31}$\\
    (1) & & & Total & 30.27$_{-2.09}^{+2.09}$ & & & 12.61$_{-1.92}^{+1.92}$ & 84.62$_{-18.94}^{+18.94}$ & 66.62$_{-18.22}^{+18.22}$\\
    (6) & Hot Ionized & Fe K & Unresolved & --- & 33000 & 0.002-0.01 & 1-6.3 & 220-1300 & 3600-21500\\
    (7) & Hot Ionized & Fe K & Unresolved & --- & 22800-42900 & 0.0029-0.0043 & 20.0 & 4400 & 80000\\
    (8) & Cold Molecular & OH & Unresolved & 1890$_{-1100}^{+2700}$ & 200-550 & 170-340 & 270$_{-130}^{+22}$ & 680$_{-290}^{+130}$ & 160$_{-70}^{+40}$\\
    (9) & Cold Molecular & CO (3-2) & Resolved & 7400 & 491 & 189 & 219 & 680 & 170\\
    (10) & Cold Molecular & CO (2-1) & Resolved & 460$_{-10}^{+10}$ & 294$_{-16}^{+16}$ & 830$_{-10}^{+10}$ & 16.6$_{-0.4}^{+0.4}$ & 31$_{-4}^{+5}$ & 9$_{-4}^{+4}$\\
    (11) & Cold Molecular & CO (1-0) & Resolved & 741 & 446 & 1052 & 32.56 & 91 & 20.4\\
    (12) & Neutral & Na I D & Resolved & 3800$_{-170}^{+270}$ & 560 & 3000 & 95$_{-6}^{+12}$ & 590$_{-40}^{+90}$ & 380$_{-30}^{+70}$\\
    (12) & Warm Ionized & [N~II] & Resolved & 230$_{-63}^{+34}$ & 423 & 3000 & 2.5$_{-0.7}^{+0.5}$ & 7.7$_{-2.1}^{+1.4}$ & 2.1$_{-0.5}^{+0.4}$\\
 \enddata
 \tablecomments{Meaning of the columns: (1) Reference, (2) Gas phase of the outflow, (3) emission line tracer used to derive the mass, (4) component of the outflow, either resolved or unresolved, (5) mass of the outflowing gas phase, (6) typical $v_{50}$ velocity of the outflow,  (7) median radius of the outflow, (8) mass outflow rate, (9) momentum outflow rate, and (10) energy outflow rate}
  \tablerefs{(1) This paper, (2) \citet{Tom2017}, (3) \citet{Rup2005}, (4) \citet{Tom2015}, (5) \citet{Vei2017}, (6) \citet{Smi2019}, (7) \citet{Nod2025}, (8) \citet{Gon2017}, (9) \citet{Flu2019}, (10) \citet{Lam2022}, (11) \citet{Lut2020}, (12) \citet{Rup2017}}
\end{deluxetable*}

% comparison of sources to each other
% comparison to other results

% both sources comparison and general statement of neon results
In the unresolved measurements of the neon gases, we measure that F11119+3257 has over twice the mass of outflowing gas ($\sim 7 \times 10^6 M_{\odot}$) as F05189-2524 ($\sim 3 \times 10^6 M_{\odot}$). When only considering mass derived from \nev\ values, our measurements are greater than all of those measured in a sample of Seyfert galaxies from the GATOS sample \citep{Zha2024}. Our ionized mass outflow rates are slightly lower than those expected from emperical relations based on the AGN luminosity from \citet{Fio2017}, but well within the observed scatter. In the breakdown between different ionization states, lower states (\neii\ and \neiii) represent the largest phases in mass, representing $\sim85\%$ of the outflow mass; however, the high velocities of the \nev\ and \nevi\ states make them a significant component of the momentum ($\sim45\%$) and energetic ($\sim65\%$) outflow rates. Based on our measurements of escape velocity for both sources discussed in Sections \ref{subsec:res_molecular} we estimate that $\sim90\%$ of the outflowing mass in both sources will escape the galaxy up to a distance of 33 R$_{\mathrm{out}}$. This is a very significant portion of the outflowing mass that will be driven to enrich the circumgalactic medium but still a relatively small component of the overall galaxy mass. Below, we compare the emission of both sources to different gas phases at varying distance scales and resolution. 

% unresolved measurements of both sources compared to other results
For F11119+3257, we can see that the warm ionized gas makes up a relatively small component of the larger-scale outflow. Compared to both the instantaneous 300 pc measurements of OH \citep[$\sim$2350 \msun;][]{Tom2015} and the time-averaged kpc scale measurements of CO(1-0) \citep[$\sim$10000 \msun;][]{Vei2017}, the mass in warm ionized gas represents $\sim3\%$ and $\sim1\%$ of the outflowing cold molecular gas. As the outflow velocities are similar between the warm ionized and cold molecular gas measurements our results have no significant impact on the large scale momentum or energy outflow rates. Previous results have shown that momentum is relatively conserved from the AGN, to the UFO, and the Kpc scale outflow \citep{Vei2017, Nar2018, Lan2024}, a conclusion unchanged by these results. Looking at the energy outflow rates, a significant component ($\sim5\%$) of the AGN luminosity drives the UFO, and $\sim30\%$ of that energy is coupled to the larger scale galactic outflow. Showing that the energy from the AGN is relatively inefficient in coupling to the host galaxy but still very significant in its impact. 

From the unresolved measurements of F05189-2524, we see a similar picture as F11119+3257: the mass in warm ionized gas in F05189-2524 makes up a relatively insignificant portion of the outflow energetics of the source. And, the resolved component represents an even smaller portion ($\sim10\%$ by mass compared to the unresolved component) of the outflow. 

Previous results show resolved measurements in cold molecular \citep{Gon2017, Flu2019, Lut2020, Lam2022}, neutral \citep{Rup2017}, and warm ionized gases \citep{Rup2017}. Results in the cold molecular gas are consistent within an order of magnitude of each other and variations between the different CO measurements are likely explained by variations in methods, instruments, and systematic errors as described in Sections 4.6.1 and 4.6.2 of \citet{Lam2022}. Regardless of the reference used, the warm ionized gas represents $\sim 0.5-5\%$ of the outflowing cold molecular gas mass. Measurements of the neutral gas \citet{Rup2017} represent another significant component of the outflow, similar in mass, momentum, and energy to the molecular gas. However, there is likely some overlap between these phases of gas. 

% discussion of results compared to NII and lack of NiII emissio
We note that our resolved warm ionized gas measurements are an order of magnitude smaller than those based on the visible \nii\ measurements in \citet{Rup2017}. This result is unlikely to be caused by any differences in metallicity or electron density measurements used in calculations. Instead, it possibly could be caused by emission line flux being attributed to nearby PAH emission or variations in our mass derivation methods compared to those used in \citet{Rup2017}.

% We attribute this difference to \nii\ probing a cooler, more mass-abundant phase of the outflow, given its lower ionization potential. MIR emission lines probing this gas phase are either too weak to observe excess emission (e.g., \arii\ and \siii) or contaminated by nearby PAH emission (\neii). We propose that this difficulty in detecting the MIR low IP emission lines in the outflow partly explains the mass discrepancy between our measurements and those in \citet{Rup2017}.

For F05189-2524, a sum of the momentum outflow rates from all of these large scale outflow phases ($\dot P_{outer}$) is consistent, given the uncertainties, with momentum conservation ($\dot P_{outer}$ $\sim$ $\dot P_{inner}$ $\sim$ $L_{bol}/c$) from the initial momentum of the AGN and UFO. The warm ionized gas represents $\leq 0.1\%$ of this momentum outflow rate. Slightly higher velocities in the warm ionized gas ($\sim$500 \kms) than the other phases cause it to have a larger impact on the energy outflow rates making up $\sim5\%$ of the total rate. Still, the overall picture is largely unchanged, as energy from the AGN is lost as distance from the center increases: $\sim5\%$ $L_{\mathrm{AGN}}$ in the sub-pc UFO and $\sim0.5\%$ $L_{\mathrm{AGN}}$ in the galactic scale kpc outflow. 

% overall impact
Overall, the addition of warm ionized gas measurements to the energetic census of both sources has no significant impact on the outflow energetics, and the conclusions from previous papers remain: the driving energetics of F11119+3257 \citep{Vei2017, Nar2018, Lan2024} and F05189-2524 \citep{Smi2019} are broadly consistent with a momentum-driven outflow. 

\subsection{Final Picture}
\label{subsec:fp}

Our final picture of these sources is remarkably similar. Both sources show clear evidence of a low-velocity rotating disk comprised of warm molecular and low-IP ionized gases. Both sources show evidence of a decrement of warm molecular gas emission in the circumnuclear region relative to warm ionized gas, evidence for radiative AGN feedback through the suppression of star formation. Both sources show evidence of a highly ionized, high-velocity spatially stratified outflow in the unresolved nuclear emission. Finally, both sources show relatively low star formation rates, supporting the AGN-driven nature of the outflow.  

The largest difference in the MIR observations of these sources comes in the detection of an extended high-velocity outflow in the warm ionized gas observations of F05189-2524, but not F11119+3257. The extent and velocity of this outflow decrease and increase, respectively, as the IP of the emission line increases. This seems to be consistent with the picture of a radiatively driven, spatially stratified outflow where higher velocities and IPs trace the region closest to the outflow axis and lower velocities trace the broader region around the edge. These extended emission maps show no evidence of a radial deceleration in the outflow.

% , decelerating outflow, but we see no significant decrease of individual line maps (Figure \ref{fig:Ion_map}) as the distance from the quasar increases. This implies that in the inner kpc, where resolved emission of all three sources is observed, the three ionization states of neon gas are moving at different velocities. 

\begin{figure*}
    \centering
    \includegraphics[width=\textwidth]{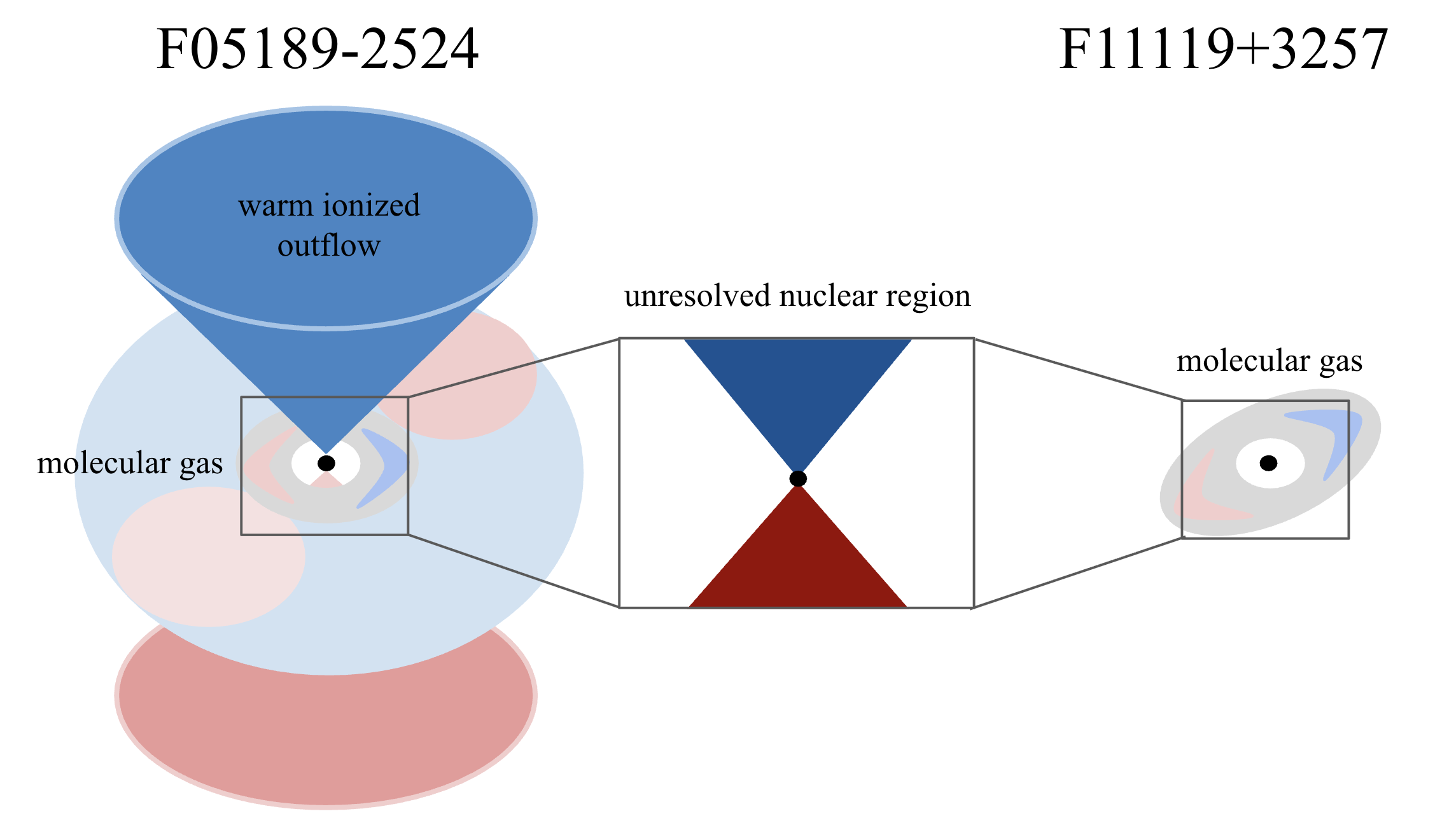}
    \caption{Schematic of the ionized outflows and surrounding disks in F05189-2514 (left) and F11119+3257 (right), derived by the \jwst\ MIR observations. Both sources show a rotating molecular disk, with the disk for F05189-2524 oriented east to west and F11119+3257 oriented southeast to northwest, going from red to blue. The strength of the molecular disk weakens significantly directly around the quasar, illustrated by the white space. F05189-2524 also exhibits a low-velocity cloud of molecular gas around the quasar. In the unresolved nuclear regions, both sources show evidence for high-velocity warm ionized outflows, but we have no indication of the direction we represent this with the inset axis. F05189-2524 also shows a biconical, warm ionized outflow extending a few kpc from the quasar with velocities slightly lower than in the unresolved region.}
    \label{fig:toymodel}
\end{figure*}

Neither source shows any evidence for outflow in the warm molecular lines. Given the significance of the CO outflow detections in these sources relative to the disk strength, we cannot rule out the possibility that a weak resolved or unresolved outflow could be present in the warm molecular gas. In F05189-2524, we do see a turbulent, lower-density clump of molecular gas to the north of the quasar in the same direction as the warm ionized outflow. This is possibly indicative of the ionized outflow stirring up the molecular gas. The only other clear evidence of interaction between the two phases is the deficit of molecular gas relative to ionized gas in the nuclear region, likely caused by AGN-driven radiative feedback.

A schematic of the geometry of the MIR emission of both of these sources is displayed in Figure \ref{fig:toymodel}. The model shows the features discussed above: low-velocity rotational disks in both sources, unresolved high-velocity outflow, and resolved low-velocity molecular gas cloud, and high-velocity ionized gas outflow in F05189-2524. 

\section{Conclusion}
\label{sec:conclusion}

We uniformly analyzed the MIRI IFS data of two strikingly similar late-stage merger ULIRGs. Both sources have shown previous evidence for strong kpc-scale winds driven by a powerful AGN-driven UFO. The new \jwst\ data provide the first high-resolution infrared view of the outflow and host galaxy in these merging galaxies. For both sources, we use \qtdfit\ to separate the quasar light from the host galaxy and analyze emission markers of dust, molecular gas, and warm ionized gas to better understand the nature of these multi-phase outflows. The main results of our analysis include: 

\begin{enumerate}    
    \item In the unresolved nuclear emission, both sources show evidence of high-velocity ($\sim 1500$ \kms) ionized gas emission indicative of a nuclear-driven, spatially stratified outflow. We measure a mass outflow rate using the neon emission lines of $10 - 14 \:\: M_{\odot}$ yr$^{-1}$ in F11119+3257 and $11-15 \:\: M_{\odot}$ yr$^{-1}$ in F05189-2524. We see no evidence for any outflowing warm molecular gas. 
    
    \item Both sources show evidence of a deficit of molecular emission ($\sim100 \times$ weaker) relative to the warm ionized gas in the nuclear region ($<1$ kpc). This deficit is roughly co-spatial with the rotating molecular and low-IP ionized gas disk and is evidence for radiative AGN feedback through suppression of molecular gas and star formation in the nuclear region of both galaxies. 

    \item In F05189-2524, we see evidence of a kpc-scale resolved high-velocity outflow in three neon emission lines \neiii, \nev, and \nevi. This emission increases in velocity and decreases in spatial extent as the IP of the line increases. The fastest emission is in the \nevi\ line, which shows a flux-weighted median outflow $|v_{50}|=1060$ \kms. The furthest extent of the emission is in the \neiii\ line extending up to $\sim2$ kpc from the central quasar. \neiii\ represents $\sim80\%$ of the mass outflow rate, the dominant component. We estimate that $\sim90\%$ of this gas exceeds the escape velocity of the galaxy. 
    
    \item The addition of the warm ionized results does not drastically alter the energetic census of these sources, typically contributing $\sim0.1-5\%$ to the overall momentum outflow rate. In comparing the momentum outflow rate from the inner UFO-driven wind to the outer multiphase outflow, both of these sources are still broadly consistent with a momentum-conserving, $\dot P_{outer}/\dot P_{inner}\sim1$, outflow. 

\end{enumerate}

%% Also note that the akcnowlodgment environment does not support long amounts of text. If you have a lot of people and institutions to acknowledge, do not use this command. Instead, create a new \section{Acknowledgments}.
\begin{acknowledgments}
K.Y.D., J.S., and S.V.\ acknowledge partial financial support by NASA for this research through STScI grants No.\  JWST-ERS-01335, JWST-GO-01865, JWST-GO-02547, JWST GO-03869, and JWST GO-05627. I.G.B. is supported by the Programa Atracci\'on de Talento Investigador ``C\'esar Nombela'' via grant 2023-T1/TEC-29030 funded by the Community of Madrid. M.P.S. acknowledges support under grants RYC2021-033094-I, CNS2023-145506, and PID2023-146667NB-I00 funded by MCIN/AEI/10.13039/501100011033 and the European Union NextGenerationEU/PRTR.
\end{acknowledgments}


\vspace{5mm}
\facilities{\jwst\ MIRI MRS}


\software{q3dfit (\citealt{Rup2014}, \citealt{Rup2021}), 
Astropy \citep{Ast2013, Ast2018}, 
matplotlib \citep{Hun2007},
NumPy \citep{Har2020},
SciPy \citep{Sci2020}
}



\appendix


\bibliography{sample631}{}
\bibliographystyle{aasjournal}



\end{document}

% End of file `sample631.tex'.